\documentclass[twocolumn,amsmath,amssymb, nobibnotes, aps, prl, longbibliography,superscriptaddress]{revtex4-2}
\usepackage{graphicx}
\usepackage{lipsum}
\usepackage{amsmath}
\usepackage{xcolor}
\usepackage{bm}
\usepackage{color}
\usepackage{amsmath,physics,bm,float}
\usepackage{soul}
\usepackage[colorlinks=true,citecolor=blue,urlcolor=blue,linkcolor=blue]{hyperref}
\usepackage{amssymb}
\usepackage{natbib}
\usepackage{upgreek}
\usepackage{mathtools}
\usepackage{siunitx}

\definecolor{myblue}{rgb}{0.125,0.125,0.75}
\usepackage{hyperref}
\hypersetup{colorlinks=true, linkcolor=myblue,urlcolor=myblue, citecolor=myblue}
\begin{document}
\title{Controlling Colloidal Flow through a Microfluidic Y-junction}
\author{Alexander P. Antonov}
\thanks{Both authors equally contributed to this work}
\affiliation{Institut f\"ur Theoretische Physik II: Weiche Materie,
Heinrich-Heine-Universit\"at D\"usseldorf, 40225 D\"usseldorf, Germany}
\author{Matthew Terkel}
\thanks{Both authors equally contributed to this work}
\affiliation{Departament de F\'{i}sica de la Mat\`{e}ria Condensada, Universitat de Barcelona, 08028 Spain}
\affiliation{Universitat de Barcelona Institute of Complex Systems, 08028 Barcelona, Spain}
\author{Fabian Jan Schwarzendahl}
\affiliation{Institut f\"ur Theoretische Physik II: Weiche Materie,
Heinrich-Heine-Universit\"at D\"usseldorf, 40225 D\"usseldorf, Germany}
\author{Carolina Rodr\'iguez-Gallo}
\affiliation{Departament de F\'{i}sica de la Mat\`{e}ria Condensada, Universitat de Barcelona, 08028 Spain}
\affiliation{Universitat de Barcelona Institute of Complex Systems, 08028 Barcelona, Spain}
\affiliation{Institute for Bioengineering of Catalonia, 08028 Barcelona, Spain}
\author{Pietro Tierno}
\email{ptierno@ub.edu}
\affiliation{Departament de F\'{i}sica de la Mat\`{e}ria Condensada, Universitat de Barcelona, 08028 Spain}
\affiliation{Universitat de Barcelona Institute of Complex Systems, 08028 Barcelona, Spain}
\author{Hartmut L\"owen}
\affiliation{Institut f\"ur Theoretische Physik II: Weiche Materie,
Heinrich-Heine-Universit\"at D\"usseldorf, 40225 D\"usseldorf, Germany}

\begin{abstract}
Microscopic particles flowing through narrow channels may accumulate near bifurcation points provoking flow reduction, 
clogging and ultimately chip breakage. Here we show that the full flow behavior of colloidal particles through 
a microfluidic Y-junction (i.e.\ a three way intersection) can be 
controlled by tuning the pair interactions and the degree of confinement. 
By combining experiments with numerical simulations, we investigate the dynamic 
states emerging when magnetizable colloids flow through a symmetric Y-junction such that a single
particle can pass through both gates with the same probability. We show that clogging can be avoided by repulsive interactions and
branching into the two channels can be steered as well by interactions: attractive particles are flowing through the same gate, 
while repulsive colloids alternate between the two gates. Even details of the particle assembly such as buckling at the exit gate are tunable by interactions and the channel geometry. 
\end{abstract}
\maketitle
Understanding the flow of particulate matter through narrow
channels is of paramount importance for many applications,
both in the natural~\cite{Hulme2008,Hofling2013,Altshuler2013,Chinappi2018,Vembadi2019} and in the synthetic~\cite{Burada2008,Bacchin2018,Reichhardt2018,Malgaretti2019} world. Examples can be found on different length
scales ranging from molecular over mesoscopic
to macroscopic sizes. Permeation of molecules
through nanopores and zeolites~\cite{Dzubiella2005} constitutes a prime
example on the molecular scale while the flow of red blood
cells in the vascular system~\cite{Patnaik1994,Yang2006,Jaggi2007} and synthetic colloids
near constrictions \cite{Genovese2011,Kreuter2013,Zimmermann2016,Sendekie2016} are topics that fall within the
colloidal regime. 
On the macroscopic scale, the
escape of pedestrians or animals through narrow gates~\cite{Zuriguel2011,Thomas2015,Tang2016},
airplane boarding~\cite{Zeineddine2017,Wittmann2019}, crowded bikers in narrow streets~\cite{Reynolds2009} and the
flow of grains through silos~\cite{Farkas2000,Zuriguel2014,Kirchner2000,Garcimart2015,Gella2017} provide other situations
known from everyday life for confined flow problems.

The transport of particles near a branching
point, such as a Y-junction, which splits the flow into two
streams of fluids is of particular  importance. Indeed, any nontrivial flow network is made
of branching points which motivates to study a
single Y-junction as a basic building block in the first
place. Key examples are biological flow networks such as the blood
circulation system~\cite{Yang2016} or synthetic 
ones~\cite{Dupire2020,Shen2023,Jorge2024}.
Moreover, Y-junctions occur on
various length scales, from nano-sized channel junctions, such as those in carbon nanotubes~\cite{Li1999}, to micron-sized artificial Y-junctions (or in extreme cases, T-junctions) in microfluidic circuits~\cite{Hymel2019,Garstecki2006,Vigolo2014,Ollila2013}
or active agents~\cite{Iyer2024}.
At the microscale, confined particles forced to pass 
through a bifurcation path may accumulate near a stagnation point and 
induce clogging, via formation 
of bridges and
arches that block the flow~\cite{Zuriguel2014}. 
This effect is responsible for the failure 
of different technological systems, spanning from microfluidic chips~\cite{Wyss2006,Dressaire2017}, to silos~\cite{Zuriguel2011}, and granular hoppers~\cite{Thomas2015,Tang2016}. Thus, understanding and controlling the emergence of clogging within simple bifurcation points such as a Y-junction is especially important.

Despite the fundamental significance of branching
points there are few experiments and simulations of particles flowing through Y-junctions.
Previous works have focused on the pure hydrodynamic
solvent flow in microfluidic junctions~\cite{Sochi2015} and also on
the transport of few particles in the flow~\cite{Ollila2013}. However,
for strongly interacting many rigid particles, such
as colloids with controllable interparticle interactions
in two-dimensional microchannel flow~\cite{Helbing2000,Koppl2006,Siems2012,Kreuter2013,Heinze2015,Foulaadvand2016}, there are no studies of the particle branching behavior
although other flow geometries such as constrictions
~\cite{Reichhardt2018,Zimmermann2016,Martens2013,Glanz2016,Hidalgo2018}, energetic barriers~\cite{Kreuter2012,Ullrich2015,Zimmermann2021}, obstacles~\cite{Borromeo2011,Ghosh2012} and shear~\cite{Nikoubashman2013,Gerloff2017} have been explored in depth
via particle-resolved simulations.

Here, we close this gap and consider the flow of strongly confined and interacting colloidal particles through a Y-junction. The
Y-junction is symmetric such that a single particle can
pass through both gates with the same probability. We
use an external field to tune the pair interaction between magnetizable paramagnetic colloids. Our finding is that the flow behavior
can be entirely tuned by the particle interactions and the geometry of the microfluidic channel. There is flow control on all levels:
the basic switch from clogging to unclogging, the partition of particles into the two different gates and even
details in the buckled structure formed by the flowing colloids. In particular, we find a tunable nonequilibrium 
branching transition between two 
distinct flowing regimes:  
a state where all particles pass through
the same gate and another one where subsequent
particles consecutively choose a different gate.

\begin{figure}[t]
\includegraphics[width=\columnwidth]{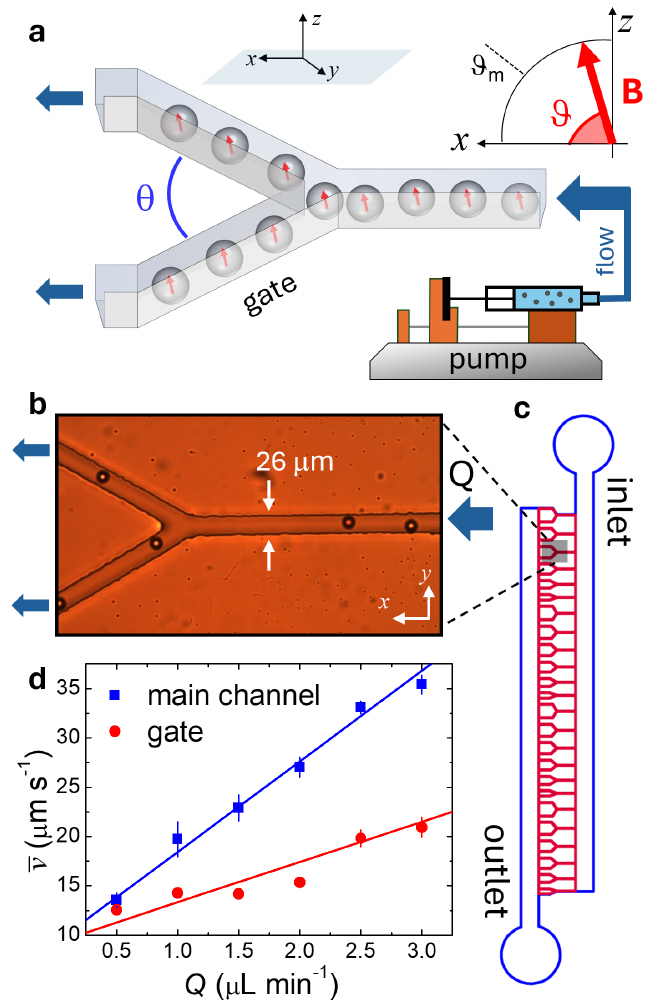}
\caption{Schematic of the experimental system: 
a syringe pump forces a collection of paramagnetic colloidal particles to flow  through a Y-junction connecting two branches at an angle $\theta=55^{\circ}$. The pair interactions between the particles are tuned by an external magnetic field $\bm{B}$
applied at an inclination angle $\vartheta \in[0,90]^{\circ}$, being $\vartheta_m=54.7^{\circ}$ the "magic angle"
delineating attractive ($\vartheta<\vartheta_m$) from repulsive ($\vartheta>\vartheta_m$) forces.
(b) Optical microscope image of one channel with flowing particles, see also MovieS1 in Supplementary Material~\cite{EPAPS}. (c) Schematic of the microfluidic chip. (d) Average particle velocity $\bar{v}$ versus flow rate $Q$ 
within the main channel (squares) and within the branches (disks). }
\label{figure1}
\end{figure}

Figs.~\ref{figure1}(a,c) show two schematics of the experimental system (a) and of the microfluidic chip (c). The latter was fabricated in polydimethylsiloxane combining UV-photolithography and microfabrication procedures, see the Supplementary Material~\cite{EPAPS} for more details. The whole chip is characterized by a series of bifurcations in form of Y-junctions connecting rectangular microfluidic channels with a constant lateral width of $d=26.3\pm 0.2 \, \rm{\mu m}$ and elevation $h\sim 100 \rm{\mu m}$, Fig.~\ref{figure1}(b).
Through these channels we disperse a water solution of 
paramagnetic polystyrene particles of diameter $\sigma=  18.8 \pm 0.4 \rm{\mu m}$ and magnetic volume susceptibility $\chi=0.014$.
Once in the channel, the particles sediment near the bottom plate
(gravity pointing along the $z$-direction) due to density mismatch, 
and display negligible thermal fluctuations due to their relative large size.
We use a syringe pump to apply a constant flow rate $Q\in [0.1,3.0] \rm{\mu L \, min^{-1}}$ which induces an average speed $\bar{v}\sim Q/A$ to each particle, being $A=d\times h$ the cross-sectional area.
Fig.~\ref{figure1}(d) shows the average velocity $\bar{v}$ measured for 
single particles both in the main channel and in one of the two branches. Since all channels have approximately the same cross-sectional area,
by flow conservation the velocity reduces by half when 
the particles enters the branches after the Y-junction.
In most of the experiments we use a small input flow such that  $\bar{v}\sim 5 \rm{\mu m s^{-1}}$ which correspond to a large P\'eclet number, $Pe=\bar{v} \sigma/(2D_0)\simeq 2000$
where the diffusion coefficient 
is negligibly small, $D_0\sim 0.01 \rm{\mu m^2 s^{-1}}$.
During transport the particles are strongly 
confined between the microfluidic walls and the bottom glass substrate,
and $\sigma=0.71 d$ so that they cannot overpass each other realizing a driven single file~\cite{Taloni2006,Barkai2009,Illien2013,Dolai2020,Cereceda2021}. 
This situation is different than previous works on flow of concentrated colloidal suspension~\cite{Isa2009,Campbell2010,Kanehl2017,Sarkar2020} where particle overtaking was possible and velocity oscillations were observed for intermediate confinement.

\begin{figure}[b]
\includegraphics[width=\columnwidth]{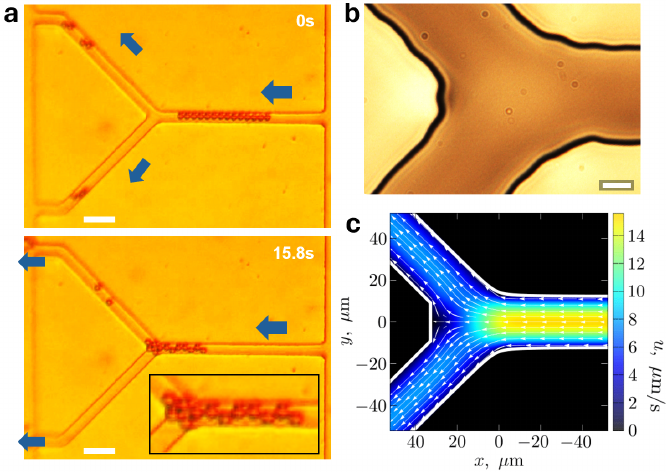} 
\caption{(a) Clogging of a chain of paramagnetic colloids within a Y-junction with $\theta=90^{\circ}$
driven at an input flow rate $Q= 0.2 \rm{\mu L min^{-1}}$. Inset shows a small enlargement of the 
clogged state, scale bar is $100 \, \rm{\mu m}$.
The corresponding video (MovieS2) can be found in~\cite{EPAPS}.
(b) Enlargement of the Y-junction section showing the 
absence of a sharp tip at the bifurcation point. Scale bar is 
$10 \, \rm{\mu m}$.
(c) Flow profile obtained by solving Eqs. \eqref{eq:navier-stokes} within a Y-junction with the lithographic imperfection.}
	\label{figure2}
\end{figure}

We now consider 
the collective transport of many particles. 
The sequence of images in Fig.~\ref{figure2}(a)
shows the typical situation encountered when an initially aligned chain of paramagnetic colloids is driven across the 
most symmetric Y-junction,  $\theta=90^{\circ}$, and in absence of any applied field. 
The particles behave as hard spheres, 
and close to the bifurcation point rather than 
choosing one of the two exit gates, 
they form a permanent clog reducing their velocity to zero
and impeding any further transport.
To understand the physical origin of this behavior, 
a close inspection of the Y-junction topology 
reveals a small but non negligible flat wedge present at the 
bifurcation point, Fig.~\ref{figure2}(b). 
This imperfection results from the resolution limit of our 
printed photolithography mask and thus, it is present in all chips investigated.  
Close to this flat wedge the fluid flow velocity $\mathbf{u} = (u_x, u_y)$ vanishes
since it generates a stagnation point. We confirm this hypothesis
by solving the incompressible Navier-Stokes equations in two dimensions
for this specific geometry with the flat edge:
\begin{subequations}
\label{eq:navier-stokes}
\begin{align}
\frac{\partial \mathbf{u}}{\partial t} + (\mathbf{u} \cdot \nabla) \mathbf{u} &= -\frac{1}{\rho_f}\nabla p + \nu \nabla^2 \mathbf{u} \\
\nabla \cdot \mathbf{u} &= 0 \, \, \, ,
\end{align}
\end{subequations}
where $p$ is the pressure, $\rho_f$ is the density of the fluid, and $\nu$ its the kinematic viscosity. 
More technical details on the boundary conditions are given in the Supplementary Material~\cite{EPAPS}.
The corresponding flow profile shown in Fig.~\ref{figure2}(c) effectively display a zero velocity region,
in contrast to the case of a sharp tip, as shown in~\cite{EPAPS}. 
The presence of this "stagnation point" can be considered as a physical obstacle that can induce clogging, and investigate the way our driven particles can overpass it.

\begin{figure}[t]
\includegraphics[width=0.9\columnwidth]{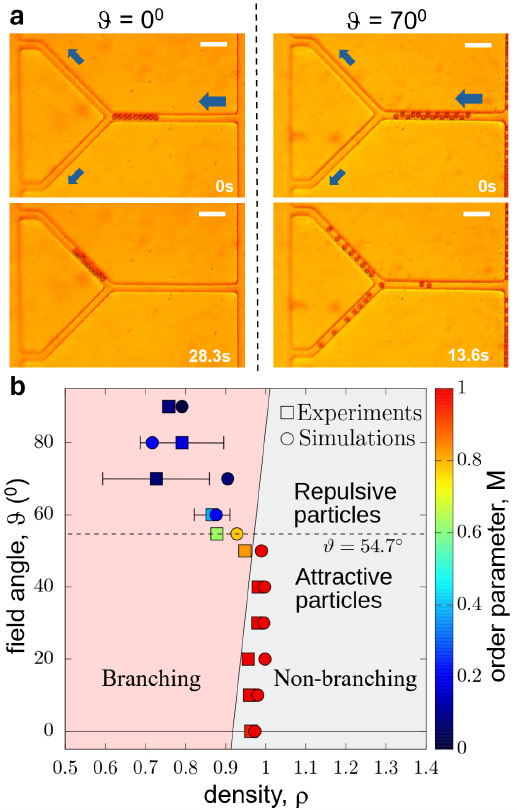}
\caption{(a) Sequence of images showing a train of paramagnetic colloidal particles flowing through a Y-junction
with $\theta=90^{\circ}$. Left (right) column illustrates the case of attractive
(repulsive) interactions induced by an in-plane field $\vartheta = 0^{\circ}$ (out of plane field, $\vartheta = 70^{\circ}$, resp.). Scale bar for all images is $100 \rm{\mu m}$, see MovieS3 in~\cite{EPAPS}.
(b) Order parameter $M$ in the $(\rho,\vartheta)$ plane, with $\rho= N\sigma/L$ denoting the normalized particle density. Scattered squares (disks) are experimental (simulation) data. The dashed line indicates $\vartheta= 54.7^{\circ}$.}
\label{figure3}
\end{figure}
\begin{figure*}[ht]
\includegraphics[width=\textwidth]{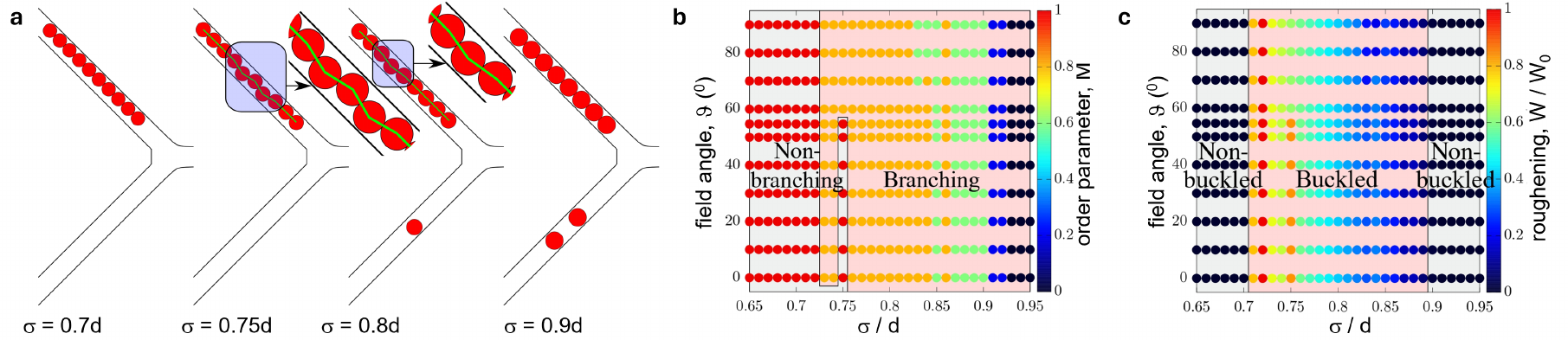}
\caption{(a) Simulation snapshots ($\vartheta=0^{\circ}$) showing the transition between the
buckled and non-buckled states and the splitting induced by varying the normalized particle size $\sigma/d$. Buckling  is highlighted in the top insets for  $\sigma = 0.75d$ and $\sigma = 0.8d$. See MovieS4 in~\cite{EPAPS}.  (b) Order parameter $M$ in the 
$(\sigma/d,\vartheta)$ plane from numerical simulations. 
(c) Normalized roughening $W/W_0$ 
of the colloidal chain in the $(\sigma/d,\vartheta)$ plane showing the buckled regions at the exit gate.}
	\label{figure4}
\end{figure*}

To unclog the system, we repeat the 
experiments and control the pair interactions 
between the particles using an 
external magnetic field $\bm{B}\equiv B_x \hat{\bm{x}}+B_z\hat{\bm{z}}$ applied along the ($\hat{\bm{x}},\hat{\bm{z}}$) plane, Fig.~\ref{figure1}(a). 
Here we fix $B = 8$ mT
and vary only the inclination angle $\vartheta$.
Under the applied field 
each particle acquires an induced dipole moment 
$\bm{m} = \pi \sigma^3 \chi \bm{B} /(6 \mu_0)$ pointing along the field direction, $\mu_0$ being the permeability of the medium (water). Thus, a pair of particles $(i,j)$ at relative position 
$r_{ij}=|\bm{r}_i-\bm{r}_j|$
experiences magnetic dipolar interactions,
\begin{equation}
U_{\rm dd}(r_{ij})=-\frac{\mu_0 m^2}{2 \pi r_{ij}^3}P_2(\cos{\vartheta})
\label{dipolar_potential} 
\end{equation} 
with $\vartheta = \tan^{-1}{(B_z/B_x)}$ the inclination angle and
$P_2(\cos{\vartheta})=(3\cos^2{\vartheta}-1)/2$ being 
the second order Legendre
polynomial. For an in-plane field ($\vartheta=0^{\circ}$) $P_2(\cos{\vartheta})=1$ dipolar interactions are attractive, and the particles within the microfluidic channel 
condense into a linear chain.
In contrast, for a perpendicular field ($\vartheta=90^{\circ}$) $P_2(\cos{\vartheta})=-1/2$ and the particle repel each other.
These interactions minimize at 
the "magic angle" $\vartheta_m=54.7^{\circ}$  
where $P_2(\cos{\vartheta_m})\sim 0$.
Thus, 
the inclination angle $\vartheta$ induces the transition between two different flowing states.  For strong attractive interactions,  all particles flow through the same gate
after crossing the bifurcation point, left column Fig.~\ref{figure3}(a).
For strong repulsion,  
the driven particles choose
alternatively a different gate, right column in Fig.~\ref{figure3}(a). 
In both cases, the flowing colloids are 
not captured by the stagnation point and system clogging is prevented.
The transition between both dynamic regimes can be 
characterized in 
terms of an effective magnetization $M\equiv |N_{\uparrow}-N_{\downarrow}|/N$ 
being $N_{\uparrow}$ ($N_{\downarrow}$) the number of particles that enters the upper (lower) branch, Fig.~\ref{figure3}(b). 
Thus, $M=1$ corresponds to the absence of splitting, which is observed in the channel before the junction and for strong, attractive interactions, $\vartheta<45^{\circ}$. 
Here the particle minimize their distance such that 
the normalized particle density $\rho =N\sigma/L \sim 1$
where $L$ is the length of the chain.
In contrast, for $\vartheta>60^{\circ}$,
the particle display initially a small transverse buckling 
just prior to reaching the Y-junction entrance with the density $\rho<1$, which decreases initially the value of 
$\rho$ as shown in the state diagram in Fig.~\ref{figure3}(b).
At the junction, the particles separate following the two gates in an alternating way, thus reducing the value of $M$ which drops to zero as all particles exit the central channel. 
The transition is smooth and crosses the magic angle where 
magnetic interactions are minimized.
Even in this situation, we find no clogging when the particles display an initial buckling state since when reaching the bifurcation point they are located along flow streamline which avoid the stagnation point, in contrast to the case of a straight chain.

We complement the experimental data with numerical simulations, as shown in Fig.~\ref{figure3}(b). In the simulations, we consider a set of $i=1,...,N$ particles, at position $\mathbf{r}_i$, which obey the 
overdamped equations of motion:
\begin{equation}
\frac{\dd \mathbf{r}_i}{\dd t} = \mathbf{u}(\mathbf{r}_i) - \frac{1}{\gamma}\nabla\left[\sum\limits_{j \ne i}U_{\rm p}(|\mathbf{r}_i - \mathbf{r}_j|) + U_{\rm w}(\mathbf{r}_i) \right] \,  .
\label{eq:dynamics}
\end{equation}
Here $\bm{u}$ is the flow field calculated from Eqs.~\ref{eq:navier-stokes}, $\gamma$ is the Stokes friction coefficient, 
$U_{\rm p}$ and $U_{\rm w}$ describe respectively 
the interaction between the particles and with the 
channel wall. The former are given via a combination of a Weeks-Chandler-Anderson (WCA) repulsive potential, $U_{\rm WCA}$, and 
the dipole-dipole interaction, $U_{\rm p}(\mathbf{r}_{ij}) = U_{\rm WCA}(\mathbf{r}_{ij}) + U_{\rm dd}(\mathbf{r}_{ij})$. More technical details are given in~\cite{EPAPS}. 

Apart from confirming the experimental 
data, the simulations allow us to set the initial particle position in the chain and to tune the particle size within the channel, 
which are two parameters difficult to control in experiments.
We observe that the lateral confinement strongly influences the dynamic states and the chain deformations via buckling in one of the two branches. In the sequence of images in Fig.~\ref{figure4}(a), we show that the branching transition $M = 1 \to M = 0$ can be also induced   by increasing the size ratio $\sigma/d$, where $M=1$ for $\sigma<0.73d$ and  $M=0$ for $\sigma>0.92d$ for all 
field inclination angles $\vartheta$. Effectively, the 
full state diagram in the $(\vartheta, \sigma/d)$ plane
in Fig.~\ref{figure4}(b) shows that this transition becomes
independent from the 
pair interactions. For strong confinement, the chain of particles enters the Y-junction
reducing its speed at contact with the flat edge (stagnation point)
and subsequently the pressure from the nearest incoming particle induces coiling and subsequent splitting between the two branches. 
In this situation excluded volume interactions dominate the final  dynamic regime. In contrast, 
smaller particles display a stronger lateral mobility 
and are able to overcome the stagnation point when initially driven 
along a streamline of the flow (Supplementary MovieS4~\cite{EPAPS}).    

Finally an intriguing consequence of the effect of confinement 
is the emergence of a buckled state within the exit gate,
as shown in the enlarged insets in Fig.~\ref{figure4}(a).
This effect can be quantified by measuring the 
chain roughening at the exit gate defined as, 
$W^2= N^{-1} \sum_{i=1}^N \langle h_i^2\rangle$, being $h_i$, $i = 1...N$ the displacement of the particle $i$ 
from the center of the corresponding branching channel after the bifurcation point, such that for  $h_i=0$ the particle 
is exactly at the channel center. 
In Fig.~\ref{figure4}(c) we show the roughness 
extracted from the particle positions by varying the size ratio $\sigma/d$
and normalized such that its maximum observed value is set to $1$. 
The corresponding diagram in the $(\sigma/d,\vartheta)$ plane shows
a re-entrant behavior where buckling is observed only for intermediate particle sizes, $\sigma/d \in [0.71,0.89]$.
For small size, the particles do not enter the stagnation point when initially located near the walls in the main channel. Therefore, they follow the streamline of the laminar flow and do not form any buckled configuration. By increasing the diameter, the particles are more likely to enter the stagnation point. However, for $\sigma/d >0.89$ the stricter confinement impedes any chain deformation.
Thus, the channel size can be used to control the colloidal assembly process at the exit gate, allowing to produce both straight or buckled structures that continuously flow along
the channel.

In conclusion, we have shown that tuning the pair interactions between strongly confined flowing particles 
can be used to avoid clogging near bifurcation points and 
to control the flow splitting into the two exit gates. 
The observed non-equilibrium branching transition at the Y-junction can be induced both by the applied field or by the confinement itself. The latter can also be used to control 
the colloidal assembly process by inducing buckling, which become absent for the extreme cases of 
large or small size ratio.

The physical situation explored with our flowing colloidal chain is also close to many biological systems. For example red-blood cells within narrow capillaries are forced to deform and proceed through narrow pores in the form of a single file~\cite{Truskey2004}. In addition, recent works have investigated the diffusive~\cite{Shen2023} and transport~\cite{Dupire2020} properties of deformable particles, as white blood cell, through an extended honeycomb network
that is formed by a series of connected Y-junctions. 
While our colloidal model system utilizes 
paramagnetic particles controlled by external fields, there are other means to tune electrostatic or steric interactions between microscopic  particles~\cite{Yethiraj2007,Han2008,Fang2020}. 
Finally, a potential future direction
could be to disperse the particles within a viscoelastic 
medium~\cite{Junot2023} to further mimic biological environments. This may lead to memory effects and 
further enrich the transition diagram unveiled by our work.  
 
M. T., C. R.-G. and P. T. 
acknowledge funding from the 
European Research Council (ERC) under the European Union's Horizon 2020 research and innovation programme (grant agreement no. 811234). P. T. acknowledges support from the Ministerio
de Ciencia e Innovaci\'on (Project No. PID2022-137713NB-C22),
the Ag\`encia de Gesti\'o d'Ajuts Universitaris i de
Recerca (Project No. 2021 SGR 00450) and the Generalitat de Catalunya under Program ``ICREA Acad\`emia''. A. A. and H. L. acknowledge funding from the DFG within project LO418/29-1.


\begin{thebibliography}{71}%
\makeatletter
\providecommand \@ifxundefined [1]{%
 \@ifx{#1\undefined}
}%
\providecommand \@ifnum [1]{%
 \ifnum #1\expandafter \@firstoftwo
 \else \expandafter \@secondoftwo
 \fi
}%
\providecommand \@ifx [1]{%
 \ifx #1\expandafter \@firstoftwo
 \else \expandafter \@secondoftwo
 \fi
}%
\providecommand \natexlab [1]{#1}%
\providecommand \enquote  [1]{``#1''}%
\providecommand \bibnamefont  [1]{#1}%
\providecommand \bibfnamefont [1]{#1}%
\providecommand \citenamefont [1]{#1}%
\providecommand \href@noop [0]{\@secondoftwo}%
\providecommand \href [0]{\begingroup \@sanitize@url \@href}%
\providecommand \@href[1]{\@@startlink{#1}\@@href}%
\providecommand \@@href[1]{\endgroup#1\@@endlink}%
\providecommand \@sanitize@url [0]{\catcode `\\12\catcode `\$12\catcode
  `\&12\catcode `\#12\catcode `\^12\catcode `\_12\catcode `\%12\relax}%
\providecommand \@@startlink[1]{}%
\providecommand \@@endlink[0]{}%
\providecommand \url  [0]{\begingroup\@sanitize@url \@url }%
\providecommand \@url [1]{\endgroup\@href {#1}{\urlprefix }}%
\providecommand \urlprefix  [0]{URL }%
\providecommand \Eprint [0]{\href }%
\providecommand \doibase [0]{https://doi.org/}%
\providecommand \selectlanguage [0]{\@gobble}%
\providecommand \bibinfo  [0]{\@secondoftwo}%
\providecommand \bibfield  [0]{\@secondoftwo}%
\providecommand \translation [1]{[#1]}%
\providecommand \BibitemOpen [0]{}%
\providecommand \bibitemStop [0]{}%
\providecommand \bibitemNoStop [0]{.\EOS\space}%
\providecommand \EOS [0]{\spacefactor3000\relax}%
\providecommand \BibitemShut  [1]{\csname bibitem#1\endcsname}%
\let\auto@bib@innerbib\@empty
\bibitem [{\citenamefont {Hulme}\ \emph {et~al.}(2008)\citenamefont {Hulme},
  \citenamefont {DiLuzio}, \citenamefont {S.~S.~Shevkoplyas}, \citenamefont
  {Mayer}, \citenamefont {Berg},\ and\ \citenamefont {Whitesides}}]{Hulme2008}%
  \BibitemOpen
  \bibfield  {author} {\bibinfo {author} {\bibfnamefont {S.~E.}\ \bibnamefont
  {Hulme}}, \bibinfo {author} {\bibfnamefont {W.~R.}\ \bibnamefont {DiLuzio}},
  \bibinfo {author} {\bibfnamefont {L.~T.}\ \bibnamefont {S.~S.~Shevkoplyas}},
  \bibinfo {author} {\bibfnamefont {M.}~\bibnamefont {Mayer}}, \bibinfo
  {author} {\bibfnamefont {H.~C.}\ \bibnamefont {Berg}},\ and\ \bibinfo
  {author} {\bibfnamefont {G.~M.}\ \bibnamefont {Whitesides}},\ }\bibfield
  {title} {\bibinfo {title} {Using ratchets and sorters to fractionate motile
  cells of escherichia coli by length},\ }\href
  {https://doi.org/10.1039/B809892A} {\bibfield  {journal} {\bibinfo  {journal}
  {Lab on a Chip}\ }\textbf {\bibinfo {volume} {8}},\ \bibinfo {pages} {1888}
  (\bibinfo {year} {2008})}\BibitemShut {NoStop}%
\bibitem [{\citenamefont {H\"ofling}\ and\ \citenamefont
  {Franosch}(2013)}]{Hofling2013}%
  \BibitemOpen
  \bibfield  {author} {\bibinfo {author} {\bibfnamefont {F.}~\bibnamefont
  {H\"ofling}}\ and\ \bibinfo {author} {\bibfnamefont {T.}~\bibnamefont
  {Franosch}},\ }\bibfield  {title} {\bibinfo {title} {Anomalous transport in
  the crowded world of biological cells},\ }\href
  {https://doi.org/10.1088/0034-4885/76/4/046602} {\bibfield  {journal}
  {\bibinfo  {journal} {Rep. Progr. Phys.}\ }\textbf {\bibinfo {volume} {76}},\
  \bibinfo {pages} {046602} (\bibinfo {year} {2013})}\BibitemShut {NoStop}%
\bibitem [{\citenamefont {Altshuler}\ \emph {et~al.}(2013)\citenamefont
  {Altshuler}, \citenamefont {Mino}, \citenamefont {P\'erez-Penichet},
  \citenamefont {R\'io}, \citenamefont {Lindner}, \citenamefont {Rousselet},\
  and\ \citenamefont {Cl\'ement}}]{Altshuler2013}%
  \BibitemOpen
  \bibfield  {author} {\bibinfo {author} {\bibfnamefont {E.}~\bibnamefont
  {Altshuler}}, \bibinfo {author} {\bibfnamefont {G.}~\bibnamefont {Mino}},
  \bibinfo {author} {\bibfnamefont {C.}~\bibnamefont {P\'erez-Penichet}},
  \bibinfo {author} {\bibfnamefont {L.~D.}\ \bibnamefont {R\'io}}, \bibinfo
  {author} {\bibfnamefont {A.}~\bibnamefont {Lindner}}, \bibinfo {author}
  {\bibfnamefont {A.}~\bibnamefont {Rousselet}},\ and\ \bibinfo {author}
  {\bibfnamefont {E.}~\bibnamefont {Cl\'ement}},\ }\bibfield  {title} {\bibinfo
  {title} {Flow-controlled densification and anomalous dispersion of e. coli
  through a constriction},\ }\href {https://doi.org/10.1039/C2SM26460A}
  {\bibfield  {journal} {\bibinfo  {journal} {Soft Matter}\ }\textbf {\bibinfo
  {volume} {9}},\ \bibinfo {pages} {1864} (\bibinfo {year} {2013})}\BibitemShut
  {NoStop}%
\bibitem [{\citenamefont {Chinappi}\ and\ \citenamefont
  {Cecconi}(2018)}]{Chinappi2018}%
  \BibitemOpen
  \bibfield  {author} {\bibinfo {author} {\bibfnamefont {M.}~\bibnamefont
  {Chinappi}}\ and\ \bibinfo {author} {\bibfnamefont {F.}~\bibnamefont
  {Cecconi}},\ }\bibfield  {title} {\bibinfo {title} {Protein sequencing via
  nanopore based devices: a nanofluidics perspective},\ }\href
  {https://doi.org/10.1088/1361-648X/aababe} {\bibfield  {journal} {\bibinfo
  {journal} {Journal of Physics: Condensed Matter}\ }\textbf {\bibinfo {volume}
  {30}},\ \bibinfo {pages} {204002} (\bibinfo {year} {2018})}\BibitemShut
  {NoStop}%
\bibitem [{\citenamefont {Vembadi}\ \emph {et~al.}(2019)\citenamefont
  {Vembadi}, \citenamefont {Menachery},\ and\ \citenamefont
  {Qasaimeh}}]{Vembadi2019}%
  \BibitemOpen
  \bibfield  {author} {\bibinfo {author} {\bibfnamefont {A.}~\bibnamefont
  {Vembadi}}, \bibinfo {author} {\bibfnamefont {A.}~\bibnamefont {Menachery}},\
  and\ \bibinfo {author} {\bibfnamefont {M.~A.}\ \bibnamefont {Qasaimeh}},\
  }\bibfield  {title} {\bibinfo {title} {Cell cytometry: Review and perspective
  on biotechnological advances},\ }\href
  {https://doi.org/10.3389/fbioe.2019.00147} {\bibfield  {journal} {\bibinfo
  {journal} {Front. Bioeng. Biotechnol.}\ }\textbf {\bibinfo {volume} {7}},\
  \bibinfo {pages} {147} (\bibinfo {year} {2019})}\BibitemShut {NoStop}%
\bibitem [{\citenamefont {Burada}\ \emph {et~al.}(2008)\citenamefont {Burada},
  \citenamefont {Schmid}, \citenamefont {Talkner}, \citenamefont {H\"anggi},
  \citenamefont {Reguera},\ and\ \citenamefont {Rub\'i}}]{Burada2008}%
  \BibitemOpen
  \bibfield  {author} {\bibinfo {author} {\bibfnamefont {P.~S.}\ \bibnamefont
  {Burada}}, \bibinfo {author} {\bibfnamefont {G.}~\bibnamefont {Schmid}},
  \bibinfo {author} {\bibfnamefont {P.}~\bibnamefont {Talkner}}, \bibinfo
  {author} {\bibfnamefont {P.}~\bibnamefont {H\"anggi}}, \bibinfo {author}
  {\bibfnamefont {D.}~\bibnamefont {Reguera}},\ and\ \bibinfo {author}
  {\bibfnamefont {J.~M.}\ \bibnamefont {Rub\'i}},\ }\bibfield  {title}
  {\bibinfo {title} {Entropic particle transport in periodic channels},\ }\href
  {https://doi.org/10.1016/j.biosystems.2008.03.006} {\bibfield  {journal}
  {\bibinfo  {journal} {Biosystems}\ }\textbf {\bibinfo {volume} {93}},\
  \bibinfo {pages} {16} (\bibinfo {year} {2008})}\BibitemShut {NoStop}%
\bibitem [{\citenamefont {Bacchin}(2018)}]{Bacchin2018}%
  \BibitemOpen
  \bibfield  {author} {\bibinfo {author} {\bibfnamefont {P.}~\bibnamefont
  {Bacchin}},\ }\bibfield  {title} {\bibinfo {title} {Interfacially driven
  transport in narrow channels},\ }\href
  {https://doi.org/10.1088/1361-648X/aacb0c} {\bibfield  {journal} {\bibinfo
  {journal} {J. Phys.: Cond. Matter}\ }\textbf {\bibinfo {volume} {39}},\
  \bibinfo {pages} {294001} (\bibinfo {year} {2018})}\BibitemShut {NoStop}%
\bibitem [{\citenamefont {Reichhardt}\ and\ \citenamefont
  {Reichhardt}(2018)}]{Reichhardt2018}%
  \BibitemOpen
  \bibfield  {author} {\bibinfo {author} {\bibfnamefont {C.~O.}\ \bibnamefont
  {Reichhardt}}\ and\ \bibinfo {author} {\bibfnamefont {C.}~\bibnamefont
  {Reichhardt}},\ }\bibfield  {title} {\bibinfo {title} {Clogging and transport
  of driven particles in asymmetric funnel arrays},\ }\href
  {https://doi.org/10.1088/1361-648X/aac247} {\bibfield  {journal} {\bibinfo
  {journal} {J. Phys. Condens. Matter}\ }\textbf {\bibinfo {volume} {30}},\
  \bibinfo {pages} {244005} (\bibinfo {year} {2018})}\BibitemShut {NoStop}%
\bibitem [{\citenamefont {Malgaretti}\ \emph {et~al.}(2019)\citenamefont
  {Malgaretti}, \citenamefont {Oshanin},\ and\ \citenamefont
  {Talbot}}]{Malgaretti2019}%
  \BibitemOpen
  \bibfield  {author} {\bibinfo {author} {\bibfnamefont {P.}~\bibnamefont
  {Malgaretti}}, \bibinfo {author} {\bibfnamefont {G.}~\bibnamefont
  {Oshanin}},\ and\ \bibinfo {author} {\bibfnamefont {J.}~\bibnamefont
  {Talbot}},\ }\bibfield  {title} {\bibinfo {title} {Special issue on transport
  in narrow channels},\ }\href {https://doi.org/10.1088/1361-648X/ab1548}
  {\bibfield  {journal} {\bibinfo  {journal} {J. Phys.: Cond. Matter}\ }\textbf
  {\bibinfo {volume} {31}},\ \bibinfo {pages} {270201} (\bibinfo {year}
  {2019})}\BibitemShut {NoStop}%
\bibitem [{\citenamefont {Dzubiella}\ and\ \citenamefont
  {Hansen}(2005)}]{Dzubiella2005}%
  \BibitemOpen
  \bibfield  {author} {\bibinfo {author} {\bibfnamefont {J.}~\bibnamefont
  {Dzubiella}}\ and\ \bibinfo {author} {\bibfnamefont {J.-P.}\ \bibnamefont
  {Hansen}},\ }\bibfield  {title} {\bibinfo {title} {Electric-field-controlled
  water and ion permeation of a hydrophobic nanopore},\ }\href
  {https://doi.org/10.1063/1.1927514} {\bibfield  {journal} {\bibinfo
  {journal} {J. Chem. Phys.}\ }\textbf {\bibinfo {volume} {122}},\ \bibinfo
  {pages} {234706} (\bibinfo {year} {2005})}\BibitemShut {NoStop}%
\bibitem [{\citenamefont {Patnaik}\ \emph {et~al.}(1994)\citenamefont
  {Patnaik}, \citenamefont {Das}, \citenamefont {Mishra}, \citenamefont
  {Mohanty}, \citenamefont {Satpathy},\ and\ \citenamefont
  {Mohanty}}]{Patnaik1994}%
  \BibitemOpen
  \bibfield  {author} {\bibinfo {author} {\bibfnamefont {J.~K.}\ \bibnamefont
  {Patnaik}}, \bibinfo {author} {\bibfnamefont {B.~S.}\ \bibnamefont {Das}},
  \bibinfo {author} {\bibfnamefont {S.~K.}\ \bibnamefont {Mishra}}, \bibinfo
  {author} {\bibfnamefont {S.}~\bibnamefont {Mohanty}}, \bibinfo {author}
  {\bibfnamefont {S.~K.}\ \bibnamefont {Satpathy}},\ and\ \bibinfo {author}
  {\bibfnamefont {D.}~\bibnamefont {Mohanty}},\ }\bibfield  {title} {\bibinfo
  {title} {Vascular clogging, mononuclear cell margination, and enhanced
  vascular permeability in the pathogenesis of human cerebral malaria},\ }\href
  {https://doi.org/10.4269/ajtmh.1994.51.642} {\bibfield  {journal} {\bibinfo
  {journal} {Am. J. Trop. Med. Hyg.}\ }\textbf {\bibinfo {volume} {3}},\
  \bibinfo {pages} {47} (\bibinfo {year} {1994})}\BibitemShut {NoStop}%
\bibitem [{\citenamefont {Yang}\ \emph {et~al.}(2006)\citenamefont {Yang},
  \citenamefont {\"Undar},\ and\ \citenamefont {Zahn}}]{Yang2006}%
  \BibitemOpen
  \bibfield  {author} {\bibinfo {author} {\bibfnamefont {S.}~\bibnamefont
  {Yang}}, \bibinfo {author} {\bibfnamefont {A.}~\bibnamefont {\"Undar}},\ and\
  \bibinfo {author} {\bibfnamefont {J.~D.}\ \bibnamefont {Zahn}},\ }\bibfield
  {title} {\bibinfo {title} {A microfluidic device for continuous, real time
  blood plasma separation},\ }\href {https://doi.org/10.1039/B516401J}
  {\bibfield  {journal} {\bibinfo  {journal} {Lab Chip}\ }\textbf {\bibinfo
  {volume} {6}},\ \bibinfo {pages} {871} (\bibinfo {year} {2006})}\BibitemShut
  {NoStop}%
\bibitem [{\citenamefont {J\"aggi}\ \emph {et~al.}(2007)\citenamefont
  {J\"aggi}, \citenamefont {Sandoz},\ and\ \citenamefont
  {Effenhauser}}]{Jaggi2007}%
  \BibitemOpen
  \bibfield  {author} {\bibinfo {author} {\bibfnamefont {R.~D.}\ \bibnamefont
  {J\"aggi}}, \bibinfo {author} {\bibfnamefont {R.}~\bibnamefont {Sandoz}},\
  and\ \bibinfo {author} {\bibfnamefont {C.~S.}\ \bibnamefont {Effenhauser}},\
  }\bibfield  {title} {\bibinfo {title} {Microfluidic depletion of red blood
  cells from whole blood in high-aspect-ratio microchannels},\ }\href
  {https://doi.org/10.1007/s10404-006-0104-9} {\bibfield  {journal} {\bibinfo
  {journal} {Microfluid Nanofluid}\ }\textbf {\bibinfo {volume} {3}},\ \bibinfo
  {pages} {47} (\bibinfo {year} {2007})}\BibitemShut {NoStop}%
\bibitem [{\citenamefont {Genovese}\ and\ \citenamefont
  {Sprakel}(2011)}]{Genovese2011}%
  \BibitemOpen
  \bibfield  {author} {\bibinfo {author} {\bibfnamefont {D.}~\bibnamefont
  {Genovese}}\ and\ \bibinfo {author} {\bibfnamefont {J.}~\bibnamefont
  {Sprakel}},\ }\bibfield  {title} {\bibinfo {title} {Crystallization and
  intermittent dynamics in constricted microfluidic flows of dense
  suspensions},\ }\href {https://doi.org/10.1039/C0SM01338B} {\bibfield
  {journal} {\bibinfo  {journal} {Soft Matter}\ }\textbf {\bibinfo {volume}
  {7}},\ \bibinfo {pages} {3889} (\bibinfo {year} {2011})}\BibitemShut
  {NoStop}%
\bibitem [{\citenamefont {Kreuter}\ \emph {et~al.}(2013)\citenamefont
  {Kreuter}, \citenamefont {Siems}, \citenamefont {Nielaba}, \citenamefont
  {Leiderer},\ and\ \citenamefont {Erbe}}]{Kreuter2013}%
  \BibitemOpen
  \bibfield  {author} {\bibinfo {author} {\bibfnamefont {C.}~\bibnamefont
  {Kreuter}}, \bibinfo {author} {\bibfnamefont {U.}~\bibnamefont {Siems}},
  \bibinfo {author} {\bibfnamefont {P.}~\bibnamefont {Nielaba}}, \bibinfo
  {author} {\bibfnamefont {P.}~\bibnamefont {Leiderer}},\ and\ \bibinfo
  {author} {\bibfnamefont {A.}~\bibnamefont {Erbe}},\ }\bibfield  {title}
  {\bibinfo {title} {Transport phenomena and dynamics of externally and
  self-propelled colloids in confined geometry},\ }\href
  {https://doi.org/10.1140/epjst/e2013-02067-x} {\bibfield  {journal} {\bibinfo
   {journal} {Eur. Phys. J. Spec. Top.}\ }\textbf {\bibinfo {volume} {222}},\
  \bibinfo {pages} {2923} (\bibinfo {year} {2013})}\BibitemShut {NoStop}%
\bibitem [{\citenamefont {Zimmermann}\ \emph {et~al.}(2016)\citenamefont
  {Zimmermann}, \citenamefont {Smallenburg},\ and\ \citenamefont
  {L\"owen}}]{Zimmermann2016}%
  \BibitemOpen
  \bibfield  {author} {\bibinfo {author} {\bibfnamefont {U.}~\bibnamefont
  {Zimmermann}}, \bibinfo {author} {\bibfnamefont {F.}~\bibnamefont
  {Smallenburg}},\ and\ \bibinfo {author} {\bibfnamefont {H.}~\bibnamefont
  {L\"owen}},\ }\bibfield  {title} {\bibinfo {title} {Flow of colloidal solids
  and fluids through constrictions: dynamical density functional theory versus
  simulation},\ }\href {https://doi.org/10.1088/0953-8984/28/24/244019}
  {\bibfield  {journal} {\bibinfo  {journal} {J. Phys.: Condens. Matter}\
  }\textbf {\bibinfo {volume} {28}},\ \bibinfo {pages} {244019} (\bibinfo
  {year} {2016})}\BibitemShut {NoStop}%
\bibitem [{\citenamefont {Sendekie}\ and\ \citenamefont
  {Bacchin}(2016)}]{Sendekie2016}%
  \BibitemOpen
  \bibfield  {author} {\bibinfo {author} {\bibfnamefont {Z.~B.}\ \bibnamefont
  {Sendekie}}\ and\ \bibinfo {author} {\bibfnamefont {P.}~\bibnamefont
  {Bacchin}},\ }\bibfield  {title} {\bibinfo {title} {Colloidal jamming
  dynamics in microchannel bottlenecks},\ }\href
  {https://doi.org/10.1021/acs.langmuir.5b04218} {\bibfield  {journal}
  {\bibinfo  {journal} {Langmuir}\ }\textbf {\bibinfo {volume} {32}},\ \bibinfo
  {pages} {1478} (\bibinfo {year} {2016})}\BibitemShut {NoStop}%
\bibitem [{\citenamefont {Zuriguel}\ \emph {et~al.}(2011)\citenamefont
  {Zuriguel}, \citenamefont {Janda}, \citenamefont {Garcimart\'{\i}n},
  \citenamefont {Lozano}, \citenamefont {Ar\'evalo},\ and\ \citenamefont
  {Maza}}]{Zuriguel2011}%
  \BibitemOpen
  \bibfield  {author} {\bibinfo {author} {\bibfnamefont {I.}~\bibnamefont
  {Zuriguel}}, \bibinfo {author} {\bibfnamefont {A.}~\bibnamefont {Janda}},
  \bibinfo {author} {\bibfnamefont {A.}~\bibnamefont {Garcimart\'{\i}n}},
  \bibinfo {author} {\bibfnamefont {C.}~\bibnamefont {Lozano}}, \bibinfo
  {author} {\bibfnamefont {R.}~\bibnamefont {Ar\'evalo}},\ and\ \bibinfo
  {author} {\bibfnamefont {D.}~\bibnamefont {Maza}},\ }\bibfield  {title}
  {\bibinfo {title} {Silo clogging reduction by the presence of an obstacle},\
  }\href {https://doi.org/10.1103/PhysRevLett.107.278001} {\bibfield  {journal}
  {\bibinfo  {journal} {Phys. Rev. Lett.}\ }\textbf {\bibinfo {volume} {107}},\
  \bibinfo {pages} {278001} (\bibinfo {year} {2011})}\BibitemShut {NoStop}%
\bibitem [{\citenamefont {Thomas}\ and\ \citenamefont
  {Durian}(2015)}]{Thomas2015}%
  \BibitemOpen
  \bibfield  {author} {\bibinfo {author} {\bibfnamefont {C.~C.}\ \bibnamefont
  {Thomas}}\ and\ \bibinfo {author} {\bibfnamefont {D.~J.}\ \bibnamefont
  {Durian}},\ }\bibfield  {title} {\bibinfo {title} {Fraction of clogging
  configurations sampled by granular hopper flow},\ }\href
  {https://doi.org/10.1103/PhysRevLett.114.178001} {\bibfield  {journal}
  {\bibinfo  {journal} {Phys. Rev. Lett.}\ }\textbf {\bibinfo {volume} {114}},\
  \bibinfo {pages} {178001} (\bibinfo {year} {2015})}\BibitemShut {NoStop}%
\bibitem [{\citenamefont {Tang}\ and\ \citenamefont
  {Behringer}(2016)}]{Tang2016}%
  \BibitemOpen
  \bibfield  {author} {\bibinfo {author} {\bibfnamefont {J.}~\bibnamefont
  {Tang}}\ and\ \bibinfo {author} {\bibfnamefont {R.~P.}\ \bibnamefont
  {Behringer}},\ }\bibfield  {title} {\bibinfo {title} {Orientation, flow, and
  clogging in a two-dimensional hopper: Ellipses vs. disks},\ }\href
  {https://doi.org/10.1209/0295-5075/114/34002} {\bibfield  {journal} {\bibinfo
   {journal} {Europhys. Lett.}\ }\textbf {\bibinfo {volume} {114}},\ \bibinfo
  {pages} {34002} (\bibinfo {year} {2016})}\BibitemShut {NoStop}%
\bibitem [{\citenamefont {Zeineddine}(2017)}]{Zeineddine2017}%
  \BibitemOpen
  \bibfield  {author} {\bibinfo {author} {\bibfnamefont {H.}~\bibnamefont
  {Zeineddine}},\ }\bibfield  {title} {\bibinfo {title} {A dynamically
  optimized aircraft boarding strategy},\ }\href
  {https://doi.org/10.1016/j.jairtraman.2016.10.010} {\bibfield  {journal}
  {\bibinfo  {journal} {J. Air Transp. Manag.}\ }\textbf {\bibinfo {volume}
  {58}},\ \bibinfo {pages} {144} (\bibinfo {year} {2017})}\BibitemShut
  {NoStop}%
\bibitem [{\citenamefont {Wittmann}(2019)}]{Wittmann2019}%
  \BibitemOpen
  \bibfield  {author} {\bibinfo {author} {\bibfnamefont {J.}~\bibnamefont
  {Wittmann}},\ }\bibfield  {title} {\bibinfo {title} {Customer-oriented
  optimization of the airplane boarding process},\ }\href
  {https://doi.org/10.1016/j.jairtraman.2019.02.002} {\bibfield  {journal}
  {\bibinfo  {journal} {J. Air Transp. Manag.}\ }\textbf {\bibinfo {volume}
  {76}},\ \bibinfo {pages} {31} (\bibinfo {year} {2019})}\BibitemShut {NoStop}%
\bibitem [{\citenamefont {Reynolds}\ \emph {et~al.}(2009)\citenamefont
  {Reynolds}, \citenamefont {Harris}, \citenamefont {Teschke}, \citenamefont
  {Cripton},\ and\ \citenamefont {Winters}}]{Reynolds2009}%
  \BibitemOpen
  \bibfield  {author} {\bibinfo {author} {\bibfnamefont {C.~C.}\ \bibnamefont
  {Reynolds}}, \bibinfo {author} {\bibfnamefont {M.~A.}\ \bibnamefont
  {Harris}}, \bibinfo {author} {\bibfnamefont {K.}~\bibnamefont {Teschke}},
  \bibinfo {author} {\bibfnamefont {P.~A.}\ \bibnamefont {Cripton}},\ and\
  \bibinfo {author} {\bibfnamefont {M.}~\bibnamefont {Winters}},\ }\bibfield
  {title} {\bibinfo {title} {The impact of transportation infrastructure on
  bicycling injuries and crashes: a review of the literature},\ }\href
  {https://doi.org/10.1186/1476-069X-8-47} {\bibfield  {journal} {\bibinfo
  {journal} {Environm. Health}\ }\textbf {\bibinfo {volume} {8}},\ \bibinfo
  {pages} {1} (\bibinfo {year} {2009})}\BibitemShut {NoStop}%
\bibitem [{\citenamefont {Helbing}\ \emph
  {et~al.}(2000{\natexlab{a}})\citenamefont {Helbing}, \citenamefont {Farkas},\
  and\ \citenamefont {Vicsek}}]{Farkas2000}%
  \BibitemOpen
  \bibfield  {author} {\bibinfo {author} {\bibfnamefont {D.}~\bibnamefont
  {Helbing}}, \bibinfo {author} {\bibfnamefont {I.}~\bibnamefont {Farkas}},\
  and\ \bibinfo {author} {\bibfnamefont {T.}~\bibnamefont {Vicsek}},\
  }\bibfield  {title} {\bibinfo {title} {Simulating dynamical features of
  escape panic},\ }\href {https://doi.org/10.1038/35035023} {\bibfield
  {journal} {\bibinfo  {journal} {Nature}\ }\textbf {\bibinfo {volume} {407}},\
  \bibinfo {pages} {487} (\bibinfo {year} {2000}{\natexlab{a}})}\BibitemShut
  {NoStop}%
\bibitem [{\citenamefont {Zuriguel}\ \emph {et~al.}(2014)\citenamefont
  {Zuriguel}, \citenamefont {Parisi}, \citenamefont {Hidalgo}, \citenamefont
  {Lozano}, \citenamefont {Janda}, \citenamefont {Gago}, \citenamefont
  {Peralta}, \citenamefont {Ferrer}, \citenamefont {Pugnaloni}, \citenamefont
  {Cl\'ement}, \citenamefont {Maza}, \citenamefont {Pagonabarraga},\ and\
  \citenamefont {Garcimart\'n}}]{Zuriguel2014}%
  \BibitemOpen
  \bibfield  {author} {\bibinfo {author} {\bibfnamefont {I.}~\bibnamefont
  {Zuriguel}}, \bibinfo {author} {\bibfnamefont {D.~R.}\ \bibnamefont
  {Parisi}}, \bibinfo {author} {\bibfnamefont {R.~C.}\ \bibnamefont {Hidalgo}},
  \bibinfo {author} {\bibfnamefont {C.}~\bibnamefont {Lozano}}, \bibinfo
  {author} {\bibfnamefont {A.}~\bibnamefont {Janda}}, \bibinfo {author}
  {\bibfnamefont {P.~A.}\ \bibnamefont {Gago}}, \bibinfo {author}
  {\bibfnamefont {J.~P.}\ \bibnamefont {Peralta}}, \bibinfo {author}
  {\bibfnamefont {L.~M.}\ \bibnamefont {Ferrer}}, \bibinfo {author}
  {\bibfnamefont {L.~A.}\ \bibnamefont {Pugnaloni}}, \bibinfo {author}
  {\bibfnamefont {E.}~\bibnamefont {Cl\'ement}}, \bibinfo {author}
  {\bibfnamefont {D.}~\bibnamefont {Maza}}, \bibinfo {author} {\bibfnamefont
  {I.}~\bibnamefont {Pagonabarraga}},\ and\ \bibinfo {author} {\bibfnamefont
  {A.}~\bibnamefont {Garcimart\'n}},\ }\bibfield  {title} {\bibinfo {title}
  {Clogging transition of many-particle systems flowing through bottlenecks},\
  }\href {https://doi.org/10.1038/srep07324} {\bibfield  {journal} {\bibinfo
  {journal} {Scientific reports}\ }\textbf {\bibinfo {volume} {4}},\ \bibinfo
  {pages} {7324} (\bibinfo {year} {2014})}\BibitemShut {NoStop}%
\bibitem [{\citenamefont {Kirchner}\ \emph {et~al.}(2003)\citenamefont
  {Kirchner}, \citenamefont {Nishinari},\ and\ \citenamefont
  {Schadschneider}}]{Kirchner2000}%
  \BibitemOpen
  \bibfield  {author} {\bibinfo {author} {\bibfnamefont {A.}~\bibnamefont
  {Kirchner}}, \bibinfo {author} {\bibfnamefont {K.}~\bibnamefont
  {Nishinari}},\ and\ \bibinfo {author} {\bibfnamefont {A.}~\bibnamefont
  {Schadschneider}},\ }\bibfield  {title} {\bibinfo {title} {Friction effects
  and clogging in a cellular automaton model for pedestrian dynamics},\ }\href
  {https://doi.org/10.1103/PhysRevE.67.056122} {\bibfield  {journal} {\bibinfo
  {journal} {Phys. Rev. E}\ }\textbf {\bibinfo {volume} {67}},\ \bibinfo
  {pages} {056122} (\bibinfo {year} {2003})}\BibitemShut {NoStop}%
\bibitem [{\citenamefont {Garcimart\'{\i}n}\ \emph {et~al.}(2015)\citenamefont
  {Garcimart\'{\i}n}, \citenamefont {Pastor}, \citenamefont {Ferrer},
  \citenamefont {Ramos}, \citenamefont {Mart\'{\i}n-G\'omez},\ and\
  \citenamefont {Zuriguel}}]{Garcimart2015}%
  \BibitemOpen
  \bibfield  {author} {\bibinfo {author} {\bibfnamefont {A.}~\bibnamefont
  {Garcimart\'{\i}n}}, \bibinfo {author} {\bibfnamefont {J.~M.}\ \bibnamefont
  {Pastor}}, \bibinfo {author} {\bibfnamefont {L.~M.}\ \bibnamefont {Ferrer}},
  \bibinfo {author} {\bibfnamefont {J.~J.}\ \bibnamefont {Ramos}}, \bibinfo
  {author} {\bibfnamefont {C.}~\bibnamefont {Mart\'{\i}n-G\'omez}},\ and\
  \bibinfo {author} {\bibfnamefont {I.}~\bibnamefont {Zuriguel}},\ }\bibfield
  {title} {\bibinfo {title} {Flow and clogging of a sheep herd passing through
  a bottleneck},\ }\href {https://doi.org/10.1103/PhysRevE.91.022808}
  {\bibfield  {journal} {\bibinfo  {journal} {Phys. Rev. E}\ }\textbf {\bibinfo
  {volume} {91}},\ \bibinfo {pages} {022808} (\bibinfo {year}
  {2015})}\BibitemShut {NoStop}%
\bibitem [{\citenamefont {Gella}\ \emph {et~al.}(2017)\citenamefont {Gella},
  \citenamefont {Maza}, \citenamefont {Zuriguel}, \citenamefont {Ashour},
  \citenamefont {Ar\'evalo},\ and\ \citenamefont {Stannarius}}]{Gella2017}%
  \BibitemOpen
  \bibfield  {author} {\bibinfo {author} {\bibfnamefont {D.}~\bibnamefont
  {Gella}}, \bibinfo {author} {\bibfnamefont {D.}~\bibnamefont {Maza}},
  \bibinfo {author} {\bibfnamefont {I.}~\bibnamefont {Zuriguel}}, \bibinfo
  {author} {\bibfnamefont {A.}~\bibnamefont {Ashour}}, \bibinfo {author}
  {\bibfnamefont {R.}~\bibnamefont {Ar\'evalo}},\ and\ \bibinfo {author}
  {\bibfnamefont {R.}~\bibnamefont {Stannarius}},\ }\bibfield  {title}
  {\bibinfo {title} {Linking bottleneck clogging with flow kinematics in
  granular materials: The role of silo width},\ }\href
  {https://doi.org/10.1103/PhysRevFluids.2.084304} {\bibfield  {journal}
  {\bibinfo  {journal} {Phys. Rev. Fluids}\ }\textbf {\bibinfo {volume} {2}},\
  \bibinfo {pages} {084304} (\bibinfo {year} {2017})}\BibitemShut {NoStop}%
\bibitem [{\citenamefont {Yang}\ \emph {et~al.}(2016)\citenamefont {Yang},
  \citenamefont {Pak},\ and\ \citenamefont {Lee}}]{Yang2016}%
  \BibitemOpen
  \bibfield  {author} {\bibinfo {author} {\bibfnamefont {J.}~\bibnamefont
  {Yang}}, \bibinfo {author} {\bibfnamefont {Y.~E.}\ \bibnamefont {Pak}},\ and\
  \bibinfo {author} {\bibfnamefont {T.~R.}\ \bibnamefont {Lee}},\ }\bibfield
  {title} {\bibinfo {title} {Predicting bifurcation angle effect on blood flow
  in the microvasculature},\ }\href {https://doi.org/10.1016/j.mvr.2016.07.001}
  {\bibfield  {journal} {\bibinfo  {journal} {Microvascular Research}\ }\textbf
  {\bibinfo {volume} {108}},\ \bibinfo {pages} {22} (\bibinfo {year}
  {2016})}\BibitemShut {NoStop}%
\bibitem [{\citenamefont {Dupire}\ \emph {et~al.}(2020)\citenamefont {Dupire},
  \citenamefont {Puech}, \citenamefont {Helfer},\ and\ \citenamefont
  {Viallat}}]{Dupire2020}%
  \BibitemOpen
  \bibfield  {author} {\bibinfo {author} {\bibfnamefont {J.}~\bibnamefont
  {Dupire}}, \bibinfo {author} {\bibfnamefont {P.-H.}\ \bibnamefont {Puech}},
  \bibinfo {author} {\bibfnamefont {E.}~\bibnamefont {Helfer}},\ and\ \bibinfo
  {author} {\bibfnamefont {A.}~\bibnamefont {Viallat}},\ }\bibfield  {title}
  {\bibinfo {title} {Mechanical adaptation of monocytes in model lung capillary
  networks},\ }\href {https://doi.org/10.1073/pnas.1919984117} {\bibfield
  {journal} {\bibinfo  {journal} {Proc. Nat. Acad. Sci. USA}\ }\textbf
  {\bibinfo {volume} {117}},\ \bibinfo {pages} {14798} (\bibinfo {year}
  {2020})}\BibitemShut {NoStop}%
\bibitem [{\citenamefont {Shen}\ \emph {et~al.}(2023)\citenamefont {Shen},
  \citenamefont {Plourabou\'e}, \citenamefont {Lintuvuori}, \citenamefont
  {Zhang}, \citenamefont {Abbasi},\ and\ \citenamefont {Misbah}}]{Shen2023}%
  \BibitemOpen
  \bibfield  {author} {\bibinfo {author} {\bibfnamefont {Z.}~\bibnamefont
  {Shen}}, \bibinfo {author} {\bibfnamefont {F.}~\bibnamefont {Plourabou\'e}},
  \bibinfo {author} {\bibfnamefont {J.~S.}\ \bibnamefont {Lintuvuori}},
  \bibinfo {author} {\bibfnamefont {H.}~\bibnamefont {Zhang}}, \bibinfo
  {author} {\bibfnamefont {M.}~\bibnamefont {Abbasi}},\ and\ \bibinfo {author}
  {\bibfnamefont {C.}~\bibnamefont {Misbah}},\ }\bibfield  {title} {\bibinfo
  {title} {Anomalous diffusion of deformable particles in a honeycomb
  network},\ }\href {https://doi.org/10.1103/PhysRevLett.130.014001} {\bibfield
   {journal} {\bibinfo  {journal} {Phys. Rev. Lett.}\ }\textbf {\bibinfo
  {volume} {130}},\ \bibinfo {pages} {014001} (\bibinfo {year}
  {2023})}\BibitemShut {NoStop}%
\bibitem [{\citenamefont {Jorge}\ \emph {et~al.}(2024)\citenamefont {Jorge},
  \citenamefont {Chardac}, \citenamefont {Poncet},\ and\ \citenamefont
  {Bartolo}}]{Jorge2024}%
  \BibitemOpen
  \bibfield  {author} {\bibinfo {author} {\bibfnamefont {C.}~\bibnamefont
  {Jorge}}, \bibinfo {author} {\bibfnamefont {A.}~\bibnamefont {Chardac}},
  \bibinfo {author} {\bibfnamefont {A.}~\bibnamefont {Poncet}},\ and\ \bibinfo
  {author} {\bibfnamefont {D.}~\bibnamefont {Bartolo}},\ }\bibfield  {title}
  {\bibinfo {title} {Active hydraulics laws from frustration principles},\
  }\href {https://doi.org/10.1038/s41567-023-02301-2} {\bibfield  {journal}
  {\bibinfo  {journal} {Nat. Phys.}\ }\textbf {\bibinfo {volume} {20}},\
  \bibinfo {pages} {303} (\bibinfo {year} {2024})}\BibitemShut {NoStop}%
\bibitem [{\citenamefont {Li}\ \emph {et~al.}(2016)\citenamefont {Li},
  \citenamefont {Papadopoulos},\ and\ \citenamefont {Xu}}]{Li1999}%
  \BibitemOpen
  \bibfield  {author} {\bibinfo {author} {\bibfnamefont {J.}~\bibnamefont
  {Li}}, \bibinfo {author} {\bibfnamefont {C.}~\bibnamefont {Papadopoulos}},\
  and\ \bibinfo {author} {\bibfnamefont {J.}~\bibnamefont {Xu}},\ }\bibfield
  {title} {\bibinfo {title} {Growing {Y}-junction carbon nanotubes},\ }\href
  {https://doi.org/10.1038/46214} {\bibfield  {journal} {\bibinfo  {journal}
  {Nature}\ }\textbf {\bibinfo {volume} {402}},\ \bibinfo {pages} {253}
  (\bibinfo {year} {2016})}\BibitemShut {NoStop}%
\bibitem [{\citenamefont {Hymel}\ \emph {et~al.}(2019)\citenamefont {Hymel},
  \citenamefont {Lan}, \citenamefont {Fujioka},\ and\ \citenamefont
  {Khismatullin}}]{Hymel2019}%
  \BibitemOpen
  \bibfield  {author} {\bibinfo {author} {\bibfnamefont {S.~J.}\ \bibnamefont
  {Hymel}}, \bibinfo {author} {\bibfnamefont {H.}~\bibnamefont {Lan}}, \bibinfo
  {author} {\bibfnamefont {H.}~\bibnamefont {Fujioka}},\ and\ \bibinfo {author}
  {\bibfnamefont {D.~B.}\ \bibnamefont {Khismatullin}},\ }\bibfield  {title}
  {\bibinfo {title} {Cell trapping in y-junction microchannels: A numerical
  study of the bifurcation angle effect in inertial microfluidics},\ }\href
  {https://doi.org/10.1063/1.5113516} {\bibfield  {journal} {\bibinfo
  {journal} {Phys. Fluids}\ }\textbf {\bibinfo {volume} {31}},\ \bibinfo
  {pages} {082003} (\bibinfo {year} {2019})}\BibitemShut {NoStop}%
\bibitem [{\citenamefont {Garstecki}\ \emph {et~al.}(2006)\citenamefont
  {Garstecki}, \citenamefont {Fuerstman}, \citenamefont {Stone},\ and\
  \citenamefont {Whitesides}}]{Garstecki2006}%
  \BibitemOpen
  \bibfield  {author} {\bibinfo {author} {\bibfnamefont {P.}~\bibnamefont
  {Garstecki}}, \bibinfo {author} {\bibfnamefont {M.~J.}\ \bibnamefont
  {Fuerstman}}, \bibinfo {author} {\bibfnamefont {H.~A.}\ \bibnamefont
  {Stone}},\ and\ \bibinfo {author} {\bibfnamefont {G.~M.}\ \bibnamefont
  {Whitesides}},\ }\bibfield  {title} {\bibinfo {title} {Formation of droplets
  and bubbles in a microfluidic t-junction-scaling and mechanism of breakup},\
  }\href {https://doi.org/10.1039/B510841A} {\bibfield  {journal} {\bibinfo
  {journal} {Lab on a Chip}\ }\textbf {\bibinfo {volume} {6}},\ \bibinfo
  {pages} {437} (\bibinfo {year} {2006})}\BibitemShut {NoStop}%
\bibitem [{\citenamefont {Vigolo}\ \emph {et~al.}(2006)\citenamefont {Vigolo},
  \citenamefont {Radl},\ and\ \citenamefont {Stone}}]{Vigolo2014}%
  \BibitemOpen
  \bibfield  {author} {\bibinfo {author} {\bibfnamefont {D.}~\bibnamefont
  {Vigolo}}, \bibinfo {author} {\bibfnamefont {S.}~\bibnamefont {Radl}},\ and\
  \bibinfo {author} {\bibfnamefont {H.~A.}\ \bibnamefont {Stone}},\ }\bibfield
  {title} {\bibinfo {title} {Unexpected trapping of particles at a t
  junction},\ }\href {https://doi.org/10.1073/pnas.1321585111} {\bibfield
  {journal} {\bibinfo  {journal} {Proc. Nat. Acad. Sci. USA}\ }\textbf
  {\bibinfo {volume} {111}},\ \bibinfo {pages} {4770} (\bibinfo {year}
  {2006})}\BibitemShut {NoStop}%
\bibitem [{\citenamefont {Ollila}\ \emph {et~al.}(2013)\citenamefont {Ollila},
  \citenamefont {Denniston},\ and\ \citenamefont {Ala-Nissila}}]{Ollila2013}%
  \BibitemOpen
  \bibfield  {author} {\bibinfo {author} {\bibfnamefont {S.~T.~T.}\
  \bibnamefont {Ollila}}, \bibinfo {author} {\bibfnamefont {C.}~\bibnamefont
  {Denniston}},\ and\ \bibinfo {author} {\bibfnamefont {T.}~\bibnamefont
  {Ala-Nissila}},\ }\bibfield  {title} {\bibinfo {title} {One- and two-particle
  dynamics in microfluidic t-junctions},\ }\href
  {https://doi.org/10.1103/PhysRevE.87.050302} {\bibfield  {journal} {\bibinfo
  {journal} {Phys. Rev. E}\ }\textbf {\bibinfo {volume} {87}},\ \bibinfo
  {pages} {050302} (\bibinfo {year} {2013})}\BibitemShut {NoStop}%
\bibitem [{\citenamefont {Iyer}\ \emph {et~al.}(2024)\citenamefont {Iyer},
  \citenamefont {Negi}, \citenamefont {Schadschneider},\ and\ \citenamefont
  {Gompper}}]{Iyer2024}%
  \BibitemOpen
  \bibfield  {author} {\bibinfo {author} {\bibfnamefont {P.}~\bibnamefont
  {Iyer}}, \bibinfo {author} {\bibfnamefont {R.~S.}\ \bibnamefont {Negi}},
  \bibinfo {author} {\bibfnamefont {A.}~\bibnamefont {Schadschneider}},\ and\
  \bibinfo {author} {\bibfnamefont {G.}~\bibnamefont {Gompper}},\ }\bibfield
  {title} {\bibinfo {title} {Interacting streams of cognitive active agents in
  a three-way intersection},\ }\href {https://arxiv.org/abs/2405.18528}
  {\bibfield  {journal} {\bibinfo  {journal} {arXiv:2405.18528}\ } (\bibinfo
  {year} {2024})}\BibitemShut {NoStop}%
\bibitem [{\citenamefont {Wyss}\ \emph {et~al.}(2006)\citenamefont {Wyss},
  \citenamefont {Blair}, \citenamefont {Morris}, \citenamefont {Stone},\ and\
  \citenamefont {Weitz}}]{Wyss2006}%
  \BibitemOpen
  \bibfield  {author} {\bibinfo {author} {\bibfnamefont {H.~M.}\ \bibnamefont
  {Wyss}}, \bibinfo {author} {\bibfnamefont {D.~L.}\ \bibnamefont {Blair}},
  \bibinfo {author} {\bibfnamefont {J.~F.}\ \bibnamefont {Morris}}, \bibinfo
  {author} {\bibfnamefont {H.~A.}\ \bibnamefont {Stone}},\ and\ \bibinfo
  {author} {\bibfnamefont {D.~A.}\ \bibnamefont {Weitz}},\ }\bibfield  {title}
  {\bibinfo {title} {Mechanism for clogging of microchannels},\ }\href
  {https://doi.org/10.1103/PhysRevE.74.061402} {\bibfield  {journal} {\bibinfo
  {journal} {Phys. Rev. E}\ }\textbf {\bibinfo {volume} {74}},\ \bibinfo
  {pages} {061402} (\bibinfo {year} {2006})}\BibitemShut {NoStop}%
\bibitem [{\citenamefont {Dressaire}\ and\ \citenamefont
  {Sauret}(2017)}]{Dressaire2017}%
  \BibitemOpen
  \bibfield  {author} {\bibinfo {author} {\bibfnamefont {E.}~\bibnamefont
  {Dressaire}}\ and\ \bibinfo {author} {\bibfnamefont {A.}~\bibnamefont
  {Sauret}},\ }\bibfield  {title} {\bibinfo {title} {Clogging of microfluidic
  systems},\ }\href {https://doi.org/10.1039/C6SM01879C} {\bibfield  {journal}
  {\bibinfo  {journal} {Soft matter}\ }\textbf {\bibinfo {volume} {13}},\
  \bibinfo {pages} {37} (\bibinfo {year} {2017})}\BibitemShut {NoStop}%
\bibitem [{\citenamefont {Sochi}(2015)}]{Sochi2015}%
  \BibitemOpen
  \bibfield  {author} {\bibinfo {author} {\bibfnamefont {T.}~\bibnamefont
  {Sochi}},\ }\bibfield  {title} {\bibinfo {title} {Fluid flow at branching
  junctions},\ }\href {https://doi.org/10.1615/InterJFluidMechRes.v42.i1.50.}
  {\bibfield  {journal} {\bibinfo  {journal} {Inter. J. Fluid Mech. Res.}\
  }\textbf {\bibinfo {volume} {42}},\ \bibinfo {pages} {59} (\bibinfo {year}
  {2015})}\BibitemShut {NoStop}%
\bibitem [{\citenamefont {Helbing}\ \emph
  {et~al.}(2000{\natexlab{b}})\citenamefont {Helbing}, \citenamefont {Farkas},\
  and\ \citenamefont {Vicsek}}]{Helbing2000}%
  \BibitemOpen
  \bibfield  {author} {\bibinfo {author} {\bibfnamefont {D.}~\bibnamefont
  {Helbing}}, \bibinfo {author} {\bibfnamefont {I.~J.}\ \bibnamefont
  {Farkas}},\ and\ \bibinfo {author} {\bibfnamefont {T.}~\bibnamefont
  {Vicsek}},\ }\bibfield  {title} {\bibinfo {title} {Freezing by heating in a
  driven mesoscopic system},\ }\href
  {https://doi.org/10.1103/PhysRevLett.84.1240} {\bibfield  {journal} {\bibinfo
   {journal} {Phys. Rev. Lett.}\ }\textbf {\bibinfo {volume} {84}},\ \bibinfo
  {pages} {1240} (\bibinfo {year} {2000}{\natexlab{b}})}\BibitemShut {NoStop}%
\bibitem [{\citenamefont {K\"oppl}\ \emph {et~al.}(2006)\citenamefont
  {K\"oppl}, \citenamefont {Henseler}, \citenamefont {Erbe}, \citenamefont
  {Nielaba},\ and\ \citenamefont {Leiderer}}]{Koppl2006}%
  \BibitemOpen
  \bibfield  {author} {\bibinfo {author} {\bibfnamefont {M.}~\bibnamefont
  {K\"oppl}}, \bibinfo {author} {\bibfnamefont {P.}~\bibnamefont {Henseler}},
  \bibinfo {author} {\bibfnamefont {A.}~\bibnamefont {Erbe}}, \bibinfo {author}
  {\bibfnamefont {P.}~\bibnamefont {Nielaba}},\ and\ \bibinfo {author}
  {\bibfnamefont {P.}~\bibnamefont {Leiderer}},\ }\bibfield  {title} {\bibinfo
  {title} {Layer reduction in driven 2d-colloidal systems through
  microchannels},\ }\href {https://doi.org/10.1103/PhysRevLett.97.208302}
  {\bibfield  {journal} {\bibinfo  {journal} {Phys. Rev. Lett.}\ }\textbf
  {\bibinfo {volume} {97}},\ \bibinfo {pages} {208302} (\bibinfo {year}
  {2006})}\BibitemShut {NoStop}%
\bibitem [{\citenamefont {Siems}\ \emph {et~al.}(2012)\citenamefont {Siems},
  \citenamefont {Kreuter}, \citenamefont {Erbe}, \citenamefont {Schwierz},
  \citenamefont {Sengupta}, \citenamefont {Leiderer},\ and\ \citenamefont
  {Nielaba}}]{Siems2012}%
  \BibitemOpen
  \bibfield  {author} {\bibinfo {author} {\bibfnamefont {U.}~\bibnamefont
  {Siems}}, \bibinfo {author} {\bibfnamefont {C.}~\bibnamefont {Kreuter}},
  \bibinfo {author} {\bibfnamefont {A.}~\bibnamefont {Erbe}}, \bibinfo {author}
  {\bibfnamefont {N.}~\bibnamefont {Schwierz}}, \bibinfo {author}
  {\bibfnamefont {S.}~\bibnamefont {Sengupta}}, \bibinfo {author}
  {\bibfnamefont {P.}~\bibnamefont {Leiderer}},\ and\ \bibinfo {author}
  {\bibfnamefont {P.}~\bibnamefont {Nielaba}},\ }\bibfield  {title} {\bibinfo
  {title} {Non-monotonic crossover from single-file to regular diffusion in
  micro-channels},\ }\href {https://doi.org/10.1038/srep01015} {\bibfield
  {journal} {\bibinfo  {journal} {Sci. Rep.}\ }\textbf {\bibinfo {volume}
  {2}},\ \bibinfo {pages} {59} (\bibinfo {year} {2012})}\BibitemShut {NoStop}%
\bibitem [{\citenamefont {Heinze}\ \emph {et~al.}(2015)\citenamefont {Heinze},
  \citenamefont {Siems},\ and\ \citenamefont {Nielaba}}]{Heinze2015}%
  \BibitemOpen
  \bibfield  {author} {\bibinfo {author} {\bibfnamefont {B.}~\bibnamefont
  {Heinze}}, \bibinfo {author} {\bibfnamefont {U.}~\bibnamefont {Siems}},\ and\
  \bibinfo {author} {\bibfnamefont {P.}~\bibnamefont {Nielaba}},\ }\bibfield
  {title} {\bibinfo {title} {Segregation of oppositely driven colloidal
  particles in hard-walled channels: A finite-size study},\ }\href
  {https://doi.org/10.1103/PhysRevE.92.012323} {\bibfield  {journal} {\bibinfo
  {journal} {Phys. Rev. E}\ }\textbf {\bibinfo {volume} {92}},\ \bibinfo
  {pages} {012323} (\bibinfo {year} {2015})}\BibitemShut {NoStop}%
\bibitem [{\citenamefont {Foulaadvand}\ and\ \citenamefont
  {Aghaee}(2016)}]{Foulaadvand2016}%
  \BibitemOpen
  \bibfield  {author} {\bibinfo {author} {\bibfnamefont {M.~E.}\ \bibnamefont
  {Foulaadvand}}\ and\ \bibinfo {author} {\bibfnamefont {B.}~\bibnamefont
  {Aghaee}},\ }\bibfield  {title} {\bibinfo {title} {Driven binary colloidal
  mixture in a 2d narrow channel with hard walls},\ }\href
  {https://doi.org/10.1140/epje/i2016-16037-2} {\bibfield  {journal} {\bibinfo
  {journal} {Eur. Phys. J. E}\ }\textbf {\bibinfo {volume} {39}},\ \bibinfo
  {pages} {37} (\bibinfo {year} {2016})}\BibitemShut {NoStop}%
\bibitem [{\citenamefont {Martens}\ \emph {et~al.}(2013)\citenamefont
  {Martens}, \citenamefont {Straube}, \citenamefont {Schmid}, \citenamefont
  {Schimansky-Geier},\ and\ \citenamefont {H\"anggi}}]{Martens2013}%
  \BibitemOpen
  \bibfield  {author} {\bibinfo {author} {\bibfnamefont {S.}~\bibnamefont
  {Martens}}, \bibinfo {author} {\bibfnamefont {A.~V.}\ \bibnamefont
  {Straube}}, \bibinfo {author} {\bibfnamefont {G.}~\bibnamefont {Schmid}},
  \bibinfo {author} {\bibfnamefont {L.}~\bibnamefont {Schimansky-Geier}},\ and\
  \bibinfo {author} {\bibfnamefont {P.}~\bibnamefont {H\"anggi}},\ }\bibfield
  {title} {\bibinfo {title} {Hydrodynamically enforced entropic trapping of
  brownian particles},\ }\href {https://doi.org/10.1103/PhysRevLett.110.010601}
  {\bibfield  {journal} {\bibinfo  {journal} {Phys. Rev. Lett.}\ }\textbf
  {\bibinfo {volume} {110}},\ \bibinfo {pages} {010601} (\bibinfo {year}
  {2013})}\BibitemShut {NoStop}%
\bibitem [{\citenamefont {Glanz}\ \emph {et~al.}(2016)\citenamefont {Glanz},
  \citenamefont {Wittkowski},\ and\ \citenamefont {L\"owen}}]{Glanz2016}%
  \BibitemOpen
  \bibfield  {author} {\bibinfo {author} {\bibfnamefont {T.}~\bibnamefont
  {Glanz}}, \bibinfo {author} {\bibfnamefont {R.}~\bibnamefont {Wittkowski}},\
  and\ \bibinfo {author} {\bibfnamefont {H.}~\bibnamefont {L\"owen}},\
  }\bibfield  {title} {\bibinfo {title} {Symmetry breaking in clogging for
  oppositely driven particles},\ }\href
  {https://doi.org/10.1103/PhysRevE.94.052606} {\bibfield  {journal} {\bibinfo
  {journal} {Phys. Rev. E}\ }\textbf {\bibinfo {volume} {94}},\ \bibinfo
  {pages} {052606} (\bibinfo {year} {2016})}\BibitemShut {NoStop}%
\bibitem [{\citenamefont {Hidalgo}\ \emph {et~al.}(2018)\citenamefont
  {Hidalgo}, \citenamefont {Go\~ni Arana}, \citenamefont {Hern\'andez-Puerta},\
  and\ \citenamefont {Pagonabarraga}}]{Hidalgo2018}%
  \BibitemOpen
  \bibfield  {author} {\bibinfo {author} {\bibfnamefont {R.~C.}\ \bibnamefont
  {Hidalgo}}, \bibinfo {author} {\bibfnamefont {A.}~\bibnamefont {Go\~ni
  Arana}}, \bibinfo {author} {\bibfnamefont {A.}~\bibnamefont
  {Hern\'andez-Puerta}},\ and\ \bibinfo {author} {\bibfnamefont
  {I.}~\bibnamefont {Pagonabarraga}},\ }\bibfield  {title} {\bibinfo {title}
  {Flow of colloidal suspensions through small orifices},\ }\href
  {https://doi.org/10.1103/PhysRevE.97.012611} {\bibfield  {journal} {\bibinfo
  {journal} {Phys. Rev. E}\ }\textbf {\bibinfo {volume} {97}},\ \bibinfo
  {pages} {012611} (\bibinfo {year} {2018})}\BibitemShut {NoStop}%
\bibitem [{\citenamefont {Kreuter}\ \emph {et~al.}(2012)\citenamefont
  {Kreuter}, \citenamefont {Siems}, \citenamefont {Henseler}, \citenamefont
  {Nielaba}, \citenamefont {Leiderer},\ and\ \citenamefont
  {Erbe}}]{Kreuter2012}%
  \BibitemOpen
  \bibfield  {author} {\bibinfo {author} {\bibfnamefont {C.}~\bibnamefont
  {Kreuter}}, \bibinfo {author} {\bibfnamefont {U.}~\bibnamefont {Siems}},
  \bibinfo {author} {\bibfnamefont {P.}~\bibnamefont {Henseler}}, \bibinfo
  {author} {\bibfnamefont {P.}~\bibnamefont {Nielaba}}, \bibinfo {author}
  {\bibfnamefont {P.}~\bibnamefont {Leiderer}},\ and\ \bibinfo {author}
  {\bibfnamefont {A.}~\bibnamefont {Erbe}},\ }\bibfield  {title} {\bibinfo
  {title} {Stochastic transport of particles across single barriers},\ }\href
  {https://doi.org/10.1088/0953-8984/24/46/464120} {\bibfield
  {journal} {\bibinfo  {journal} {J. Phys.: Condens. Matter}\ }\textbf
  {\bibinfo {volume} {24}},\ \bibinfo {pages} {464120} (\bibinfo {year}
  {2012})}\BibitemShut {NoStop}%
\bibitem [{\citenamefont {Siems}\ and\ \citenamefont
  {Nielaba}(2015)}]{Ullrich2015}%
  \BibitemOpen
  \bibfield  {author} {\bibinfo {author} {\bibfnamefont {U.}~\bibnamefont
  {Siems}}\ and\ \bibinfo {author} {\bibfnamefont {P.}~\bibnamefont
  {Nielaba}},\ }\bibfield  {title} {\bibinfo {title} {Transport and diffusion
  properties of interacting colloidal particles in two-dimensional
  microchannels with a periodic potential},\ }\href
  {https://doi.org/10.1103/PhysRevE.91.022313} {\bibfield  {journal} {\bibinfo
  {journal} {Phys. Rev. E}\ }\textbf {\bibinfo {volume} {91}},\ \bibinfo
  {pages} {022313} (\bibinfo {year} {2015})}\BibitemShut {NoStop}%
\bibitem [{\citenamefont {Zimmermann}\ \emph {et~al.}(2021)\citenamefont
  {Zimmermann}, \citenamefont {L\"owen}, \citenamefont {Kreuter}, \citenamefont
  {Erbe}, \citenamefont {Leiderer},\ and\ \citenamefont
  {Smallenburg}}]{Zimmermann2021}%
  \BibitemOpen
  \bibfield  {author} {\bibinfo {author} {\bibfnamefont {U.}~\bibnamefont
  {Zimmermann}}, \bibinfo {author} {\bibfnamefont {H.}~\bibnamefont {L\"owen}},
  \bibinfo {author} {\bibfnamefont {C.}~\bibnamefont {Kreuter}}, \bibinfo
  {author} {\bibfnamefont {A.}~\bibnamefont {Erbe}}, \bibinfo {author}
  {\bibfnamefont {P.}~\bibnamefont {Leiderer}},\ and\ \bibinfo {author}
  {\bibfnamefont {F.}~\bibnamefont {Smallenburg}},\ }\bibfield  {title}
  {\bibinfo {title} {Negative resistance for colloids driven over two barriers
  in a microchannel},\ }\href {https://doi.org/10.1039/D0SM01700K} {\bibfield
  {journal} {\bibinfo  {journal} {Soft Matter}\ }\textbf {\bibinfo {volume}
  {17}},\ \bibinfo {pages} {516} (\bibinfo {year} {2021})}\BibitemShut
  {NoStop}%
\bibitem [{\citenamefont {Borromeo}\ \emph {et~al.}(2011)\citenamefont
  {Borromeo}, \citenamefont {Marchesoni},\ and\ \citenamefont
  {Ghosh}}]{Borromeo2011}%
  \BibitemOpen
  \bibfield  {author} {\bibinfo {author} {\bibfnamefont {M.}~\bibnamefont
  {Borromeo}}, \bibinfo {author} {\bibfnamefont {F.}~\bibnamefont
  {Marchesoni}},\ and\ \bibinfo {author} {\bibfnamefont {P.}~\bibnamefont
  {Ghosh}},\ }\bibfield  {title} {\bibinfo {title} {Communication: Driven
  brownian transport in eccentric septate channels},\ }\href
  {https://doi.org/10.1063/1.3535559} {\bibfield  {journal} {\bibinfo
  {journal} {J. Chem. Phys.}\ }\textbf {\bibinfo {volume} {134}},\ \bibinfo
  {pages} {051101} (\bibinfo {year} {2011})}\BibitemShut {NoStop}%
\bibitem [{\citenamefont {Ghosh}\ \emph {et~al.}(2012)\citenamefont {Ghosh},
  \citenamefont {H\"anggi}, \citenamefont {Marchesoni}, \citenamefont
  {Martens}, \citenamefont {Nori}, \citenamefont {Schimansky-Geier},\ and\
  \citenamefont {Schmid}}]{Ghosh2012}%
  \BibitemOpen
  \bibfield  {author} {\bibinfo {author} {\bibfnamefont {P.~K.}\ \bibnamefont
  {Ghosh}}, \bibinfo {author} {\bibfnamefont {P.}~\bibnamefont {H\"anggi}},
  \bibinfo {author} {\bibfnamefont {F.}~\bibnamefont {Marchesoni}}, \bibinfo
  {author} {\bibfnamefont {S.}~\bibnamefont {Martens}}, \bibinfo {author}
  {\bibfnamefont {F.}~\bibnamefont {Nori}}, \bibinfo {author} {\bibfnamefont
  {L.}~\bibnamefont {Schimansky-Geier}},\ and\ \bibinfo {author} {\bibfnamefont
  {G.}~\bibnamefont {Schmid}},\ }\bibfield  {title} {\bibinfo {title} {Driven
  brownian transport through arrays of symmetric obstacles},\ }\href
  {https://doi.org/10.1103/PhysRevE.85.011101} {\bibfield  {journal} {\bibinfo
  {journal} {Phys. Rev. E}\ }\textbf {\bibinfo {volume} {85}},\ \bibinfo
  {pages} {011101} (\bibinfo {year} {2012})}\BibitemShut {NoStop}%
\bibitem [{\citenamefont {Nikoubashman}\ \emph {et~al.}(2013)\citenamefont
  {Nikoubashman}, \citenamefont {Likos},\ and\ \citenamefont
  {Kahl}}]{Nikoubashman2013}%
  \BibitemOpen
  \bibfield  {author} {\bibinfo {author} {\bibfnamefont {A.}~\bibnamefont
  {Nikoubashman}}, \bibinfo {author} {\bibfnamefont {C.~N.}\ \bibnamefont
  {Likos}},\ and\ \bibinfo {author} {\bibfnamefont {G.}~\bibnamefont {Kahl}},\
  }\bibfield  {title} {\bibinfo {title} {Computer simulations of colloidal
  particles under flow in microfluidic channels},\ }\href
  {https://doi.org/10.1039/C2SM26727F} {\bibfield  {journal} {\bibinfo
  {journal} {Soft Matter}\ }\textbf {\bibinfo {volume} {9}},\ \bibinfo {pages}
  {2603} (\bibinfo {year} {2013})}\BibitemShut {NoStop}%
\bibitem [{\citenamefont {Gerloff}\ \emph {et~al.}(2017)\citenamefont
  {Gerloff}, \citenamefont {Vezirov},\ and\ \citenamefont
  {Klapp}}]{Gerloff2017}%
  \BibitemOpen
  \bibfield  {author} {\bibinfo {author} {\bibfnamefont {S.}~\bibnamefont
  {Gerloff}}, \bibinfo {author} {\bibfnamefont {T.~A.}\ \bibnamefont
  {Vezirov}},\ and\ \bibinfo {author} {\bibfnamefont {S.~H.~L.}\ \bibnamefont
  {Klapp}},\ }\bibfield  {title} {\bibinfo {title} {Shear-induced laning
  transition in a confined colloidal film},\ }\href
  {https://doi.org/10.1103/PhysRevE.95.062605} {\bibfield  {journal} {\bibinfo
  {journal} {Phys. Rev. E}\ }\textbf {\bibinfo {volume} {95}},\ \bibinfo
  {pages} {062605} (\bibinfo {year} {2017})}\BibitemShut {NoStop}%
\bibitem [{EPA()}]{EPAPS}%
  \BibitemOpen
  \href@noop {} {}\bibinfo {note} {See Document at https:// which includes
  four supplementary videos and text to support the main results.}\BibitemShut
  {Stop}%
\bibitem [{\citenamefont {Taloni}\ and\ \citenamefont
  {Marchesoni}(2006)}]{Taloni2006}%
  \BibitemOpen
  \bibfield  {author} {\bibinfo {author} {\bibfnamefont {A.}~\bibnamefont
  {Taloni}}\ and\ \bibinfo {author} {\bibfnamefont {F.}~\bibnamefont
  {Marchesoni}},\ }\bibfield  {title} {\bibinfo {title} {Single-file diffusion
  on a periodic substrate},\ }\href
  {https://doi.org/10.1103/PhysRevLett.96.020601} {\bibfield  {journal}
  {\bibinfo  {journal} {Phys. Rev. Lett.}\ }\textbf {\bibinfo {volume} {96}},\
  \bibinfo {pages} {020601} (\bibinfo {year} {2006})}\BibitemShut {NoStop}%
\bibitem [{\citenamefont {Barkai}\ and\ \citenamefont
  {Silbey}(2009)}]{Barkai2009}%
  \BibitemOpen
  \bibfield  {author} {\bibinfo {author} {\bibfnamefont {E.}~\bibnamefont
  {Barkai}}\ and\ \bibinfo {author} {\bibfnamefont {R.}~\bibnamefont
  {Silbey}},\ }\bibfield  {title} {\bibinfo {title} {Theory of single file
  diffusion in a force field},\ }\href
  {https://doi.org/10.1103/PhysRevLett.102.050602} {\bibfield  {journal}
  {\bibinfo  {journal} {Phys. Rev. Lett.}\ }\textbf {\bibinfo {volume} {102}},\
  \bibinfo {pages} {050602} (\bibinfo {year} {2009})}\BibitemShut {NoStop}%
\bibitem [{\citenamefont {Illien}\ \emph {et~al.}(2013)\citenamefont {Illien},
  \citenamefont {B\'enichou}, \citenamefont {Mej\'{\i}a-Monasterio},
  \citenamefont {Oshanin},\ and\ \citenamefont {Voituriez}}]{Illien2013}%
  \BibitemOpen
  \bibfield  {author} {\bibinfo {author} {\bibfnamefont {P.}~\bibnamefont
  {Illien}}, \bibinfo {author} {\bibfnamefont {O.}~\bibnamefont {B\'enichou}},
  \bibinfo {author} {\bibfnamefont {C.}~\bibnamefont {Mej\'{\i}a-Monasterio}},
  \bibinfo {author} {\bibfnamefont {G.}~\bibnamefont {Oshanin}},\ and\ \bibinfo
  {author} {\bibfnamefont {R.}~\bibnamefont {Voituriez}},\ }\bibfield  {title}
  {\bibinfo {title} {Active transport in dense diffusive single-file systems},\
  }\href {https://doi.org/10.1103/PhysRevLett.111.038102} {\bibfield  {journal}
  {\bibinfo  {journal} {Phys. Rev. Lett.}\ }\textbf {\bibinfo {volume} {111}},\
  \bibinfo {pages} {038102} (\bibinfo {year} {2013})}\BibitemShut {NoStop}%
\bibitem [{\citenamefont {Dolai}\ \emph {et~al.}(2020)\citenamefont {Dolai},
  \citenamefont {Das}, \citenamefont {Kundu}, \citenamefont {Dasgupta},
  \citenamefont {Dhar},\ and\ \citenamefont {Kumar}}]{Dolai2020}%
  \BibitemOpen
  \bibfield  {author} {\bibinfo {author} {\bibfnamefont {P.}~\bibnamefont
  {Dolai}}, \bibinfo {author} {\bibfnamefont {A.}~\bibnamefont {Das}}, \bibinfo
  {author} {\bibfnamefont {A.}~\bibnamefont {Kundu}}, \bibinfo {author}
  {\bibfnamefont {C.}~\bibnamefont {Dasgupta}}, \bibinfo {author}
  {\bibfnamefont {A.}~\bibnamefont {Dhar}},\ and\ \bibinfo {author}
  {\bibfnamefont {K.~V.}\ \bibnamefont {Kumar}},\ }\bibfield  {title} {\bibinfo
  {title} {Universal scaling in active single-file dynamics},\ }\href
  {https://doi.org/10.1039/D0SM00687D} {\bibfield  {journal} {\bibinfo
  {journal} {Soft Matter}\ }\textbf {\bibinfo {volume} {16}},\ \bibinfo {pages}
  {7077} (\bibinfo {year} {2020})}\BibitemShut {NoStop}%
\bibitem [{\citenamefont {Cereceda-L\'opez}\ \emph {et~al.}(2021)\citenamefont
  {Cereceda-L\'opez}, \citenamefont {Lips}, \citenamefont {Ortiz-Ambriz},
  \citenamefont {Ryabov}, \citenamefont {Maass},\ and\ \citenamefont
  {Tierno}}]{Cereceda2021}%
  \BibitemOpen
  \bibfield  {author} {\bibinfo {author} {\bibfnamefont {E.}~\bibnamefont
  {Cereceda-L\'opez}}, \bibinfo {author} {\bibfnamefont {D.}~\bibnamefont
  {Lips}}, \bibinfo {author} {\bibfnamefont {A.}~\bibnamefont {Ortiz-Ambriz}},
  \bibinfo {author} {\bibfnamefont {A.}~\bibnamefont {Ryabov}}, \bibinfo
  {author} {\bibfnamefont {P.}~\bibnamefont {Maass}},\ and\ \bibinfo {author}
  {\bibfnamefont {P.}~\bibnamefont {Tierno}},\ }\bibfield  {title} {\bibinfo
  {title} {Hydrodynamic interactions can induce jamming in flow-driven
  systems},\ }\href {https://doi.org/10.1103/PhysRevLett.127.214501} {\bibfield
   {journal} {\bibinfo  {journal} {Phys. Rev. Lett.}\ }\textbf {\bibinfo
  {volume} {127}},\ \bibinfo {pages} {214501} (\bibinfo {year}
  {2021})}\BibitemShut {NoStop}%
\bibitem [{\citenamefont {Isa}\ \emph {et~al.}(2009)\citenamefont {Isa},
  \citenamefont {Besseling}, \citenamefont {Morozov},\ and\ \citenamefont
  {Poon}}]{Isa2009}%
  \BibitemOpen
  \bibfield  {author} {\bibinfo {author} {\bibfnamefont {L.}~\bibnamefont
  {Isa}}, \bibinfo {author} {\bibfnamefont {R.}~\bibnamefont {Besseling}},
  \bibinfo {author} {\bibfnamefont {A.~N.}\ \bibnamefont {Morozov}},\ and\
  \bibinfo {author} {\bibfnamefont {W.~C.~K.}\ \bibnamefont {Poon}},\
  }\bibfield  {title} {\bibinfo {title} {Velocity oscillations in microfluidic
  flows of concentrated colloidal suspensions},\ }\href
  {https://doi.org/10.1103/PhysRevLett.102.058302} {\bibfield  {journal}
  {\bibinfo  {journal} {Phys. Rev. Lett.}\ }\textbf {\bibinfo {volume} {102}},\
  \bibinfo {pages} {058302} (\bibinfo {year} {2009})}\BibitemShut {NoStop}%
\bibitem [{\citenamefont {Campbell}\ and\ \citenamefont
  {Haw}(2010)}]{Campbell2010}%
  \BibitemOpen
  \bibfield  {author} {\bibinfo {author} {\bibfnamefont {A.~I.}\ \bibnamefont
  {Campbell}}\ and\ \bibinfo {author} {\bibfnamefont {M.~D.}\ \bibnamefont
  {Haw}},\ }\bibfield  {title} {\bibinfo {title} {Jamming and unjamming of
  concentrated colloidal dispersions in channel flows},\ }\href
  {https://doi.org/10.1039/C0SM00110D} {\bibfield  {journal} {\bibinfo
  {journal} {Soft Matter}\ }\textbf {\bibinfo {volume} {6}},\ \bibinfo {pages}
  {4688} (\bibinfo {year} {2010})}\BibitemShut {NoStop}%
\bibitem [{\citenamefont {Kanehl}\ and\ \citenamefont
  {Stark}(2017)}]{Kanehl2017}%
  \BibitemOpen
  \bibfield  {author} {\bibinfo {author} {\bibfnamefont {P.}~\bibnamefont
  {Kanehl}}\ and\ \bibinfo {author} {\bibfnamefont {H.}~\bibnamefont {Stark}},\
  }\bibfield  {title} {\bibinfo {title} {Self-organized velocity pulses of
  dense colloidal suspensions in microchannel flow},\ }\href
  {https://doi.org/10.1103/PhysRevLett.119.018002} {\bibfield  {journal}
  {\bibinfo  {journal} {Phys. Rev. Lett.}\ }\textbf {\bibinfo {volume} {119}},\
  \bibinfo {pages} {018002} (\bibinfo {year} {2017})}\BibitemShut {NoStop}%
\bibitem [{\citenamefont {Sarkar}\ \emph {et~al.}(2020)\citenamefont {Sarkar},
  \citenamefont {Chaudhuri},\ and\ \citenamefont {Sain}}]{Sarkar2020}%
  \BibitemOpen
  \bibfield  {author} {\bibinfo {author} {\bibfnamefont {T.}~\bibnamefont
  {Sarkar}}, \bibinfo {author} {\bibfnamefont {P.}~\bibnamefont {Chaudhuri}},\
  and\ \bibinfo {author} {\bibfnamefont {A.}~\bibnamefont {Sain}},\ }\bibfield
  {title} {\bibinfo {title} {Poiseuille flow of soft polycrystals in 2d rough
  channels},\ }\href {https://doi.org/10.1103/PhysRevLett.124.158003}
  {\bibfield  {journal} {\bibinfo  {journal} {Phys. Rev. Lett.}\ }\textbf
  {\bibinfo {volume} {124}},\ \bibinfo {pages} {158003} (\bibinfo {year}
  {2020})}\BibitemShut {NoStop}%
\bibitem [{\citenamefont {Truskey}\ \emph {et~al.}(2004)\citenamefont
  {Truskey}, \citenamefont {Yuan},\ and\ \citenamefont {Katz}}]{Truskey2004}%
  \BibitemOpen
  \bibfield  {author} {\bibinfo {author} {\bibfnamefont {G.~A.}\ \bibnamefont
  {Truskey}}, \bibinfo {author} {\bibfnamefont {F.}~\bibnamefont {Yuan}},\ and\
  \bibinfo {author} {\bibfnamefont {D.~F.}\ \bibnamefont {Katz}},\ }\href@noop
  {} {\emph {\bibinfo {title} {Transport Phenomena in Biological System}}}\
  (\bibinfo  {publisher} {Pearson/Prentice Hall},\ \bibinfo {address} {Upper
  Saddle River, N.J.},\ \bibinfo {year} {2004})\BibitemShut {NoStop}%
\bibitem [{\citenamefont {Yethiraj}(2007)}]{Yethiraj2007}%
  \BibitemOpen
  \bibfield  {author} {\bibinfo {author} {\bibfnamefont {A.}~\bibnamefont
  {Yethiraj}},\ }\bibfield  {title} {\bibinfo {title} {Tunable colloids:
  control of colloidal phase transitions with tunable interactions},\ }\href
  {https://doi.org/10.1039/B704251P} {\bibfield  {journal} {\bibinfo  {journal}
  {Soft Matter}\ }\textbf {\bibinfo {volume} {3}},\ \bibinfo {pages} {1099}
  (\bibinfo {year} {2007})}\BibitemShut {NoStop}%
\bibitem [{\citenamefont {Han}\ \emph {et~al.}(2008)\citenamefont {Han},
  \citenamefont {Shokef}, \citenamefont {Alsayed}, \citenamefont {Yunker},
  \citenamefont {Lubensky},\ and\ \citenamefont {Yodh}}]{Han2008}%
  \BibitemOpen
  \bibfield  {author} {\bibinfo {author} {\bibfnamefont {Y.}~\bibnamefont
  {Han}}, \bibinfo {author} {\bibfnamefont {Y.}~\bibnamefont {Shokef}},
  \bibinfo {author} {\bibfnamefont {A.~M.}\ \bibnamefont {Alsayed}}, \bibinfo
  {author} {\bibfnamefont {P.}~\bibnamefont {Yunker}}, \bibinfo {author}
  {\bibfnamefont {T.~C.}\ \bibnamefont {Lubensky}},\ and\ \bibinfo {author}
  {\bibfnamefont {A.~G.}\ \bibnamefont {Yodh}},\ }\bibfield  {title} {\bibinfo
  {title} {Geometric frustration in buckled colloidal monolayers},\ }\href
  {https://doi.org/10.1038/nature07595} {\bibfield  {journal} {\bibinfo
  {journal} {Nature}\ }\textbf {\bibinfo {volume} {456}},\ \bibinfo {pages}
  {898} (\bibinfo {year} {2008})}\BibitemShut {NoStop}%
\bibitem [{\citenamefont {Fang}\ \emph {et~al.}(2020)\citenamefont {Fang},
  \citenamefont {Hagan},\ and\ \citenamefont {Rogers}}]{Fang2020}%
  \BibitemOpen
  \bibfield  {author} {\bibinfo {author} {\bibfnamefont {H.}~\bibnamefont
  {Fang}}, \bibinfo {author} {\bibfnamefont {M.}~\bibnamefont {Hagan}},\ and\
  \bibinfo {author} {\bibfnamefont {W.}~\bibnamefont {Rogers}},\ }\bibfield
  {title} {\bibinfo {title} {Two-step crystallization and solid-solid
  transitions in binary colloidal mixtures},\ }\href
  {https://doi.org/10.1073/pnas.2008561117} {\bibfield  {journal} {\bibinfo
  {journal} {Proc. Natl. Acad. Sci. USA}\ }\textbf {\bibinfo {volume} {117}},\
  \bibinfo {pages} {27927} (\bibinfo {year} {2020})}\BibitemShut {NoStop}%
\bibitem [{\citenamefont {Junot}\ \emph {et~al.}(2023)\citenamefont {Junot},
  \citenamefont {De~Corato},\ and\ \citenamefont {Tierno}}]{Junot2023}%
  \BibitemOpen
  \bibfield  {author} {\bibinfo {author} {\bibfnamefont {G.}~\bibnamefont
  {Junot}}, \bibinfo {author} {\bibfnamefont {M.}~\bibnamefont {De~Corato}},\
  and\ \bibinfo {author} {\bibfnamefont {P.}~\bibnamefont {Tierno}},\
  }\bibfield  {title} {\bibinfo {title} {Large scale zigzag pattern emerging
  from circulating active shakers},\ }\href
  {https://doi.org/10.1103/PhysRevLett.131.068301} {\bibfield  {journal}
  {\bibinfo  {journal} {Phys. Rev. Lett.}\ }\textbf {\bibinfo {volume} {131}},\
  \bibinfo {pages} {068301} (\bibinfo {year} {2023})}\BibitemShut {NoStop}%
\end{thebibliography}
\end{document}